\documentclass[showpacs,aps,prl,twocolumn,superscriptaddress]{revtex4-1}
\usepackage[english]{babel}
\usepackage{graphicx}
\usepackage{color}
\usepackage{xcolor}
\usepackage{amsmath}
\usepackage{amssymb}
\usepackage[pdftex,breaklinks]{hyperref}

\usepackage{hyperref}
\usepackage{epsfig}
\usepackage{color}
\usepackage{psfrag}
\usepackage{float}
\usepackage{tikz}
\usepackage{caption}
\usepackage{subcaption}
\usepackage{amsmath}
\usepackage{amssymb}

\newcommand{\rem}[1]{}

\begin{document}

\title{
Chirality and dimensionality in the ultrastrong light-matter coupling regime
}
\author{R.~Avriller}
\affiliation{Univ.~Bordeaux, CNRS, LOMA, UMR 5798, F-33405 Talence, France}
\email{remi.avriller@u-bordeaux.fr}
\author{C.~Genet}
\affiliation{University of Strasbourg and CNRS, CESQ and ISIS, UMR 7006, F-67000 Strasbourg, France}
%

\begin{abstract}
We unveil the key-role of dimensionality in describing chiroptical properties of molecules embedded inside an optical Fabry-P\'erot cavity. 
For a 2D-layer configuration, we show that the interplay between molecular chirality and spatial dispersion of the cavity-modes, results in a gyrotropic coupling $\chi$ at the origin of a differential shift in polaritonic energy-spectra.
This differential shift is proportional to $\chi$, while for 3D bulk-aggregate configurations it is shown to vanish.  
We interpret physically the former 2D-chiral effect by analogy with the classical Newtonian motion of a fictive particle in presence of 3D restoring force, and static magnetic field. 
The gyrotropic coupling is shown to directly perturbate the anholonomy angle of the classical trajectories, and the fictive particle undergoes cyclotron gyrations upon entering the ultrastrong light-matter coupling regime.
\end{abstract}

\date{\today}

\maketitle

%
Developing a better understanding of chiral light-matter interactions is an important issue, in order to design more precise spectroscopic tools for molecular \cite{yang2011vibrational}, chemical \cite{domingos2018sensing,gottarelli2008use} and biological \cite{ranjbar2009circular,miles2021tools} sciences.
This quest is however made difficult due to the intrinsically weak intensity of chiroptical signals \cite{RevModPhys.9.432}, and to the complexity and multiscale nature of the related light-matter interaction process. 
Indeed, at the microscopic level, a description of optical activity (OA) or circular dichroism (CD) in three dimensions (3D) involves a delicate interference process between intrinsic molecular electronic or vibrational properties, and the spatial inhomogeneity of the local electromagnetic (EM) field \cite{RevModPhys.9.432,landau2013electrodynamics,lindell1994electromagnetic,thirunamachandran1998molecular}. 
A great amount of efforts was thus devoted to develop enantioselective schemes, in the linear and ultrafast optical regimes \cite{10.1063/1.4930109,D2CP01009G,PhysRevX.13.011044}.
In parallel, chemical engineering approaches have been proposed to optimize molecular chiroptical responses, scaling them to larger supramolecular ensembles \cite{crassous2009chiral,seeber2006supramolecular}.   
At a larger scale, optical properties of a macroscopic chiral object, result from another interference process between the EM fields scattered by the many elementary scattering centers making the object \cite{fernandez-corbaton_objects_2016}.
Such is the case of patterned two-dimensional (2D) chiral metasurfaces \cite{fedotov_asymmetric_2006,solomon_enantiospecific_2019,plum_chiral_2015,semnani2020spin,menzel_advanced_2010,oh_chiral_2015}, that are designed to imprint a specific signature of their intrinsic geometric chirality into the polarization-dependence of their scattered EM field.  
It thus appears necessary to unveil symmetry properties of the system "chiral material plus electromagnetic-field" as a whole \cite{fernandez-corbaton_objects_2016,andrews2018quantum}, and to describe the dominant role played by geometry and dimensionality of the scatterer \cite{gautier_planar_2022} for properly describing chiral light-matter interactions.
Electromagnetic Fabry-P\'erot (FP) cavities are ideal plateforms to probe those issues \cite{hubener2021engineering}.
Indeed, collective interaction mechanisms \cite{10.1021/acs.accounts.6b00295,doi:10.1126/science.abd0336} can easily take place in such cavities between electronic \cite{https://doi.org/10.1002/anie.201107033} or vibrational molecular transitions \cite{george_multiple_2016} and quantum vacuum fluctuations of the cavity-modes.
In the strong-coupling (SC) regime, both material responses and vacuum fields hybridize, with spectral polaritonic features characterized by a collective vacuum Rabi spliting (cVRS). When the cVRS is of the same order as the resonant cavity-mode frequency, the coupled system enters the ultrastrong coupling (USC) regime \cite{ciuti_quantum_2005}. In both regimes, alterations of physical \cite{10.1063/PT.3.4749,hirai2023molecular,delpo2021polariton,appugliese2022breakdown} and chemical \cite{https://doi.org/10.1002/anie.201107033,https://doi.org/10.1002/anie.201605504,doi:10.1126/science.ade7147} properties of the cavity-embedded materials have been reported.
It thus appears natural to ask whether chiroptical properties can also be modified under similar coupling conditions.
Unfortunately, chiroptical interactions inside the cavity are dominated by polarization reversal of the EM waves at the interfaces with the metallic mirrors \cite{jackson2021classical}, leading to no clear amplification of OA or CD-signals \cite{plum_chiral_2015,  voronin_single-handedness_2022,scott_enhanced_2020}, despite a very large cVRS \cite{mauro_classical_2024}.
The idea was thus proposed to recycle the EM-field helicity in the cavity either by using spin-preserving mirrors instead of normal metallic ones \cite{yoo_chiral_2015,feis_helicity-preserving_2020,voronin_single-handedness_2022,baranov_toward_2023,mauro_chiral_2023,mauro_classical_2024,PhysRevLett.132.043602}, or by introducing a chiral seed inside cavity \cite{gautier_planar_2022,chen20242d,salij2024theory} to induce an imbalance in polarization content of the cavity-modes.
Recent theoretical works investigated some aspects of the interaction mechanism developing in such chiral cavities, either by solving classical Maxwell equations but with a poor microscopic description of the material properties \cite{voronin_single-handedness_2022,mauro_chiral_2023,mauro_classical_2024}, or by using mixed quantum electrodynamics approaches \cite{thirunamachandran1998molecular,PhysRevA.25.2473} with ab-initio quantum chemical methods \cite{rokaj2018light,schafer2020relevance,riso_strong_2023} to describe accurately the molecular electronic and vibrational structure, but at the price of truncating the number of cavity-modes to one or few symmetry-broken chiral modes \cite{schafer_chiral_2023,riso_strong_2023,riso2023strong}.
None of those approaches are completely satisfactory, and a general understanding is still missing regarding the tight relation existing between the information encoded into the polarization or spatial structure of the cavity EM-field that is of topological nature \cite{bliokh_spinorbit_2015,bian2024angular}, and the multiscale interaction mechanism between this field and the embedded chiral molecules. 
%

%
We have realized however that the USC regime inside a FP cavity forms a perfect setting to make explicit the above mentioned missing tight relation. Exploiting the unique features of USC, we show in this Letter that, even in absence of chiral mirrors, the interplay between a lower dimensionality of the coupled molecular structure and spatial dispersion of the EM cavity-modes is sufficient to generate a mechanism of gyrotropic coupling. Importantly, this collective, dimensional gyrotropic coupling is induced even in the absence of chiral mirrors, a result that immediately opens relevant perspectives in the context of current polaritonic physics and chemistry. 
We analyse in depth the implications of our theoretical model. One key implication is a new relation between dimensionality and topology of the associated classical trajectories. 
This relation sheds a new light on collective chiral interactions developing in the USC regime within an achiral FP cavity.
%

\begin{figure}[tb]
\includegraphics[width=1.0\linewidth, angle=-0]{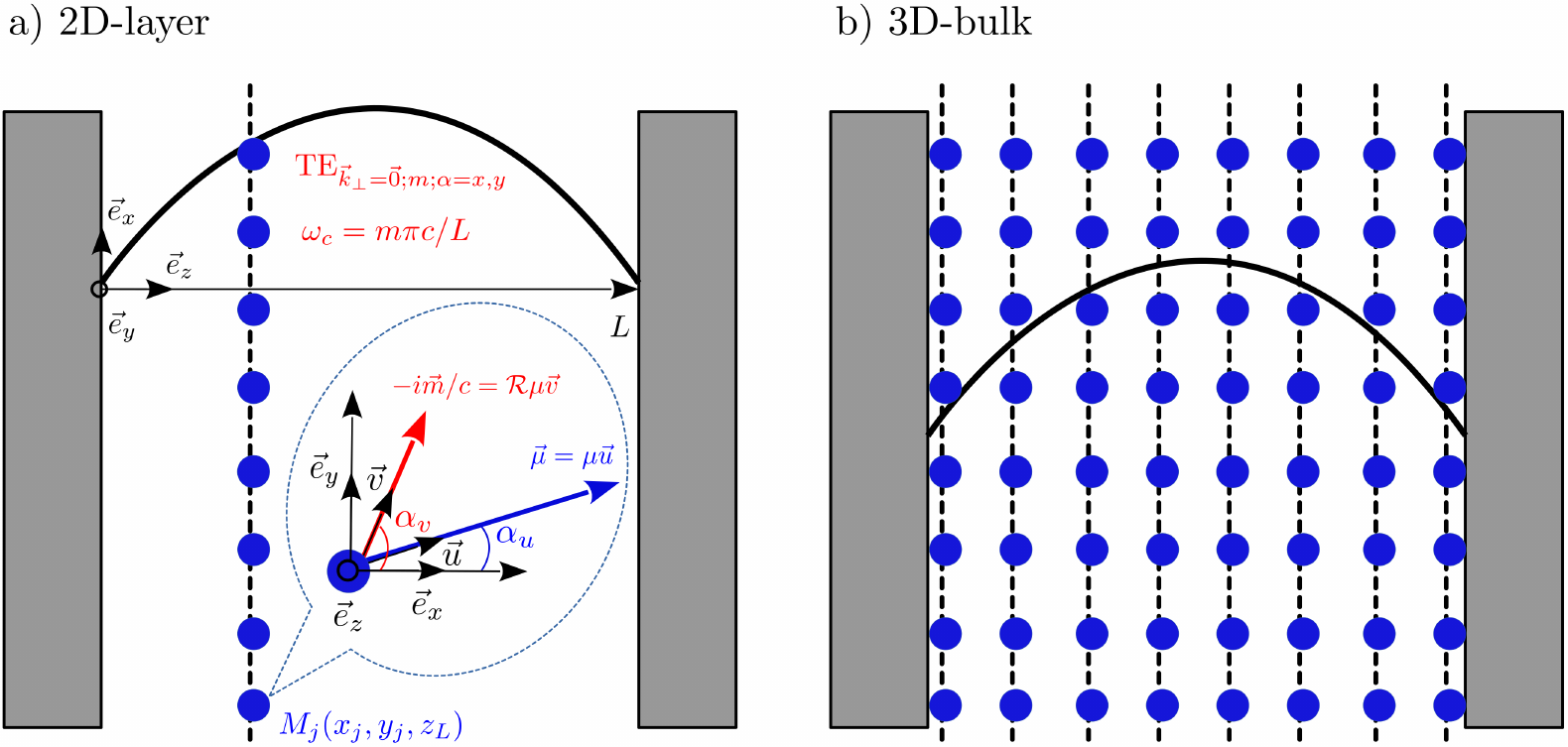}
\caption{
a) Scheme of optical Fabry-P\'erot cavity of length $L$.
A 2D-layer of molecules is located at position $z=z_L$, and couples to two $\mbox{TE}_{\vec{k}_\perp=\vec{0};m=1;\alpha=x,y}$ cavity-modes.
Inset: Zoom of molecule $M_j$ (blue dot) with electric (blue) and magnetic (red) transition-dipole $\vec{\mu}$ and $\vec{m}$ respectively. 
b) Same scheme but for a 3D-bulk configuration of molecules filling homogeneously the cavity. 
\label{fig:Fig1}}
\end{figure}

%
%
\textit{Cavity set-up.---} The above mentioned theoretical model is the most efficient and simple one, for which the optical Fabry-P\'erot cavity is made of two perfectly reflecting metallic mirrors, located at a distance $L$ one from each other (see Fig.~\ref{fig:Fig1}).
In the following, we label $z$ the cavity optical axis, $x$,$y$ its orthogonal transverse directions, and $\vec{u}_\perp$ the component of any $\vec{u}$-vector projected onto the $(x,y)$ plane.
The microscopic interaction mechanism at the origin of polariton-formation is generated microscopically by molecules having an electronic transition-frequency $\Delta$ between their ground and excited state, which is resonant or close to resonant with some cavity-mode of frequency $\omega_c=m\pi L/c$, with $m\in \mathbb{N}^*$ the mode index and $c$ the speed of light.
In order to deal with the interaction of an ensemble of chiral molecules inside an achiral FP cavity, it appears necessary to include a modelling of gyrotropic interactions developing between the embedded molecules and at least two cavity-modes, here two transverse-electric (TE) optical modes $\mbox{TE}_{\vec{k}_\perp=\vec{0};m;\alpha}$ with transverse k-vector $\vec{k}_\perp=\vec{0}$ and $\alpha=x,y$ polarization. 
We probe the impact of dimensionality with two different cavity-configurations: i) a mesoscopic 2D-layer located at position $z_L$ (see Fig.~\ref{fig:Fig1}-a)), and ii) a macroscopic 3D-bulk aggregate (see Fig.~\ref{fig:Fig1}-b)) homogeneously dispersed inside the cavity. 
We suppose that each molecules $M_j$ has the same electric and magnetic transition-dipoles $\vec{\mu} = \mu \vec{u} \in \mathbb{R}^3$ and $\vec{m}/c = i \mathcal{R}\mu \vec{v}\in i\mathbb{R}^3$ (see inset of Fig.~\ref{fig:Fig1}-a)), which points in a direction given respectively by the transverse unit-vectors $\vec{u} = \cos\left( \alpha_u \right) \vec{e}_x +  \sin\left( \alpha_u \right) \vec{e}_y$ and $\vec{v}=\cos\left( \alpha_v \right) \vec{e}_x +  \sin\left( \alpha_v \right) \vec{e}_y$, with $\alpha_v = \alpha_u + \alpha$. 
We write $\mu$ the matrix-element of the electric transition-dipole, and $\mathcal{R}$ a dimensionless ratio characterizing the magnetic transition-dipole.
The molecule is optically active when the quantity (known as the molecular rotational strength) $\mbox{Im}\left(\vec{\mu} \cdot \vec{m}^*\right) \ne 0$ \cite{thirunamachandran1998molecular,RevModPhys.9.432}, the sign of which characterizes the molecular enantiomeric class.
%
%
%
Each $M_j$ molecule couples to the local electromagnetic field-operator (see Supplementary material \cite{SupMat}), giving rise to a cVRS frequency \cite{10.1021/acs.accounts.6b00295,10.1063/1.881201} 
\begin{eqnarray}
\Omega_R &=& \frac{1}{\hbar} \sqrt{\frac{\hbar\omega_c N}{\varepsilon_0 V}}\mbox{ }\mu
\label{Vacuum_Rabi_Splitting} \, , 
\end{eqnarray}
with $\hbar$ the Planck constant, $\varepsilon_0$ the vacuum permittivity, $N$ the total number of coupled molecules and $V$ the cavity mode-volume.
For molecules located in the layer $z=z_L$, the additional cross-coupling between the transition dipoles $\vec{\mu}$ and $\vec{m}$ and the cavity-modes, gives rise to a gyrotropic coupling \cite{SupMat} quantified by the parameter
\begin{eqnarray}
\chi({z_L}) &=& \mathcal{R}\left\lbrack \vec{e}_z \cdot\left( \vec{u} \times \vec{v} \right)\right\rbrack\sin\left( 2\theta_{z_L} \right)
\label{Gyrotropic_Coupling} \, , 
\end{eqnarray}
with the angle $\theta_{z_L}=m\pi z_L/L$. 
This fundamental dimensionless quantity is a true-scalar (with respect to both rotations and planar reflections), resulting from the hybridization between chiral molecules and the cavity modes: it is the product of one term proportional to $\mathcal{R}$ times the projection of $ \vec{u} \times \vec{v}$ on the $z$-direction characterizing the molecular orientation and enantiomeric class, by a second term $\sin\left( 2\theta_{z_L} \right)$ associated to the spatial dispersion of the TE-modes along the $z$-axis.    
The $\chi({z_L})$-parameter depends crucially on the interplay between the dimensionality of the molecular ensemble and the spatial dispersion of the cavity-modes.
It can change sign, either by changing the molecular orientation, the molecular enantiomeric class, or by changing the molecular position $z_L$ with respect to a spatial plane of symmetry of the cavity-mode profile.
We anticipate a shift of polaritonic-spectra due to the $\chi({z_L})$-parameter when the molecules are prepared in a 2D-layer configuration located at $z=z_L$, while the corresponding 3D-bulk configuration results in its vanishing after spatial $z$-integration of all layers along the optical axis.  
%

%
\textit{Hopfield polaritonic spectra.---} 
%
%
\begin{figure}[tb]
\includegraphics[width=1.0\linewidth, angle=-0]{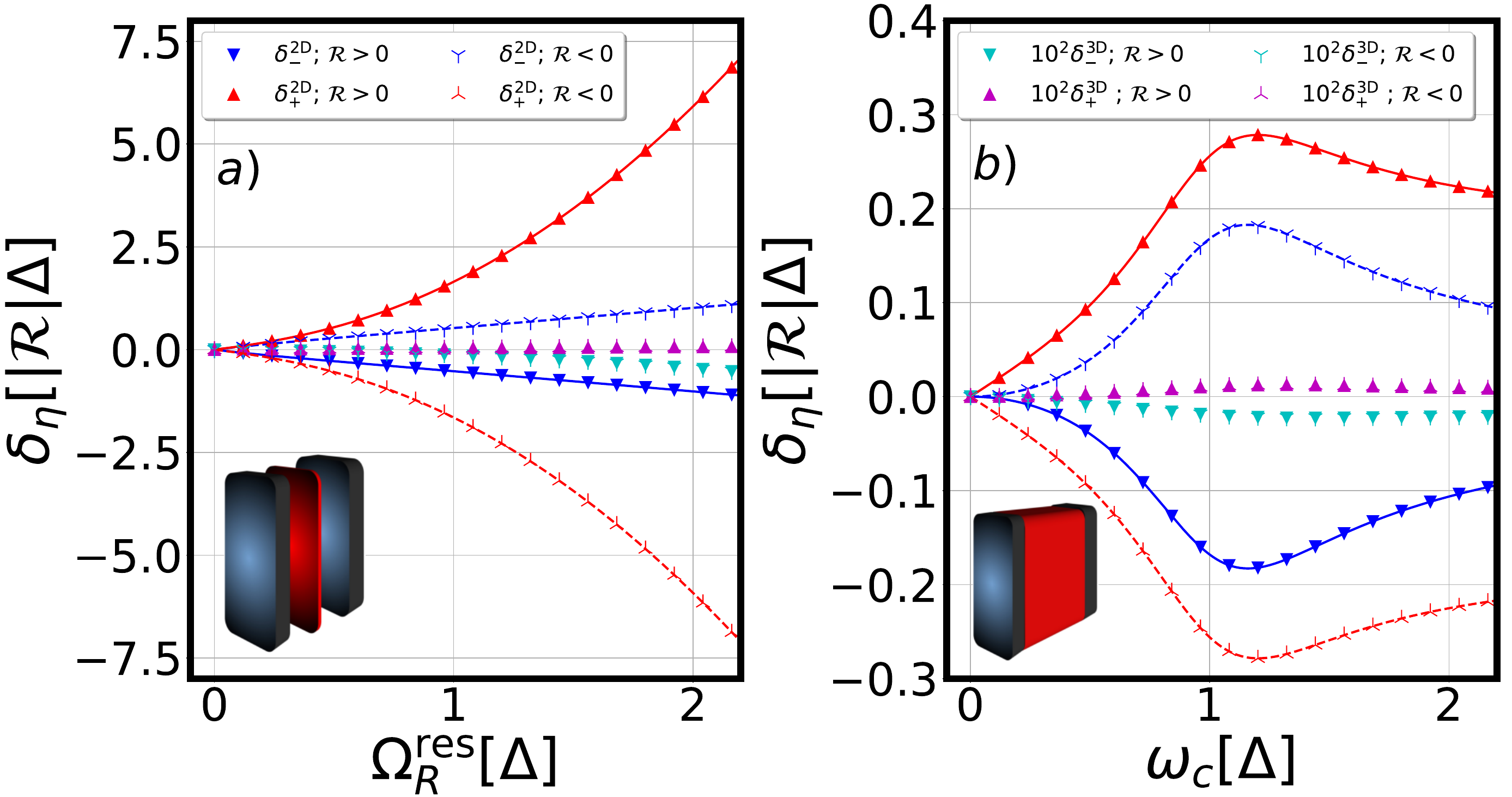}
\caption{
a) Plots of the relative shift in the polaritonic spectra $\delta_\eta = \omega_{\eta;\mathcal{R}} - \omega_{\eta;\mathcal{R}=0}$ (in units of $\left\lvert \mathcal{R} \right\rvert\Delta$) computed at resonance ($\omega_c = \tilde{\Delta}$), as a function of the collective vacuum Rabi-splitting $\Omega^{\rm{res}}_R$.
The upper (lower) triangles and crosses are the spectra of the upper (lower) polariton $\eta=1 (-1)$, computed after numerical diagonalization of the Hopfield matrix.
Triangles (crosses) are obtained for molecules with $\mathcal{R} = 10^{-3} (-10^{-3})$.
Plain and dashed curves are the analytical results of Eq.~\ref{delta_spectra_1} for the 2D-layer.
The 3D-bulk case is multiplied by a factor $10^2$.
b) Same plots, but as a function of cavity frequency $\omega_c$ at fixed $\Omega^{\rm{res}}_R=0.3$.  
Parameters common to both panels: $\Delta=1.0$, $m=1$, $z_L=\frac{L}{4}$, $\alpha_u=\frac{\pi}{4}$, $\alpha=\frac{\pi}{2}$, and $\chi({z_L})=\mathcal{R}$.
\label{fig:Fig2}}
\end{figure}
We further investigate the above mentioned differential shift in polaritonic frequency spectra $\delta_{\eta} = \omega_{\eta;\mathcal{R}} - \omega_{\eta;\mathcal{R}=0}$, with $\omega_{\eta;\mathcal{R}}$ the polariton frequency for the $\eta=+(-)$ upper (lower) polariton at finite value of $\mathcal{R}$. 
For the 2D-layer (see inset of Fig.~\ref{fig:Fig2}-a)), we derived an analytical formula for $\delta^{\rm{2D}}_\eta$ (see \cite{SupMat}), that is valid in the limit $\mathcal{R} \ll 1$ (fulfilled in practice) 
\begin{eqnarray}
\delta^{\rm{2D}}_\eta &\approx & 
\eta \chi({z_L}) \left( \frac{\Omega_R}{\tilde{\Omega}}\right)^2  \omega_{\eta;0} 
\label{delta_spectra_1} \, , 
\end{eqnarray}
where we introduced
\begin{eqnarray}
\tilde{\Omega}^2 &=& \sqrt{ 
\left( \frac{\omega^2_c - \tilde{\Delta}^2}{2} \right)^2 
+
4 \Delta \omega_c \Omega^2_R \left\lVert \vec{u}_\perp \right\rVert^2 \sin^2\left(\theta_z\right) } 
\label{delta_spectra_2} \, , 
\end{eqnarray}
with $\tilde{\Delta}=\sqrt{\Delta^2 + 4\frac{\Delta}{\omega_c}\Omega^2_R \left\lVert \vec{u}_\perp \right\rVert^2\sin^2\left( \theta_z\right)}$, and the polariton frequency $\omega_{\eta;0}  =  \sqrt{
\frac{\omega^2_c + \tilde{\Delta}^2}{2} 
+ \eta \tilde{\Omega}^2 }$ for $\mathcal{R}=0$.
We show in Fig.~\ref{fig:Fig2}-a), the $\delta^{\rm{2D}}_{\eta}$-signals computed after exact numerical diagonalization of the bozonized Hopfield Hamiltonian (see \cite{SupMat}) describing the 2D-layer configuration, as a function of the cVRS $\Omega^{\rm{res}}_R = \Omega_R$.
Same curves are shown in Fig.~\ref{fig:Fig2}-b), but as a function of cavity frequency $\omega_c$.
A very good agreement is obtained between the analytical result in Eq.~\ref{delta_spectra_1} (plain and dashed curves) and the exact numerical solution (triangles and crosses).
We find that, the  upper polariton $\delta^{\rm{2D}}_+$-signal is positive for $\chi({z_L})>0$ (upper red triangles) and negative for $\chi({z_L})<0$ (upper red crosses).
The reversed signs are obtained for the lower polariton $\delta^{\rm{2D}}_-$-signals (blue curves). 
Consistently with Eqs.~\ref{Gyrotropic_Coupling},\ref{delta_spectra_1}, the computed signals exhibit the remarkable antisymmetry property $\delta^{\rm{2D}}_\eta(-\mathcal{R}) = -\delta^{\rm{2D}}_\eta(\mathcal{R})$ upon sign-reversal of $\mathcal{R}$, on the full range of probed parameters. 
The latter enantiodependent relation is the main signature of the reduced dimensionality of the cavity configuration on the considered chiroptical observables.
We note that for low values of the cVRS ($\Omega_R \ll \omega_c$), Eq.~\ref{delta_spectra_1} predicts that $\delta^{\rm{2D}}_\eta \approx \frac{\chi({z_L})}{2} \left( \eta \frac{\Omega_R}{\left\lVert \vec{u}_\perp \right\rVert} + \frac{\Omega^2_R}{\Delta}\right)$.  
The latter signal is thus approximately linear with $\Omega_R$, consistently with the collective nature of the polaritons, which leads to $\delta^{\rm{2D}}_-(\mathcal{R}) \approx - \delta^{\rm{2D}}_+(\mathcal{R})$.
However, departures do occur at larger cVRS (first quadrically) in Fig.~\ref{fig:Fig2}-a), that are more pronounced for the lower than for the upper polariton. 
In the developed USC regime, no such antisymmetry relation exists between the $\delta^{\rm{2D}}_+$ and $\delta^{\rm{2D}}_-$-signals, which in general are not opposite $\delta^{\rm{2D}}_-(\mathcal{R}) \ne - \delta^{\rm{2D}}_+(\mathcal{R})$.
We assign the latter inequality to the contribution of both counter-rotating and self-dipoles interaction terms \cite{rokaj2018light,schafer2020relevance} in the microscopic Hamiltonian (see \cite{SupMat}), which similarly to the case of polaritonic gaps \cite{ciuti_quantum_2005,george_multiple_2016}, provide here asymmetric or ``chiral polaritonic band-gaps'' fingerprints upon entering the USC regime. 
%
%
\begin{figure}[tb]
\includegraphics[width=1.0\linewidth, angle=-0]{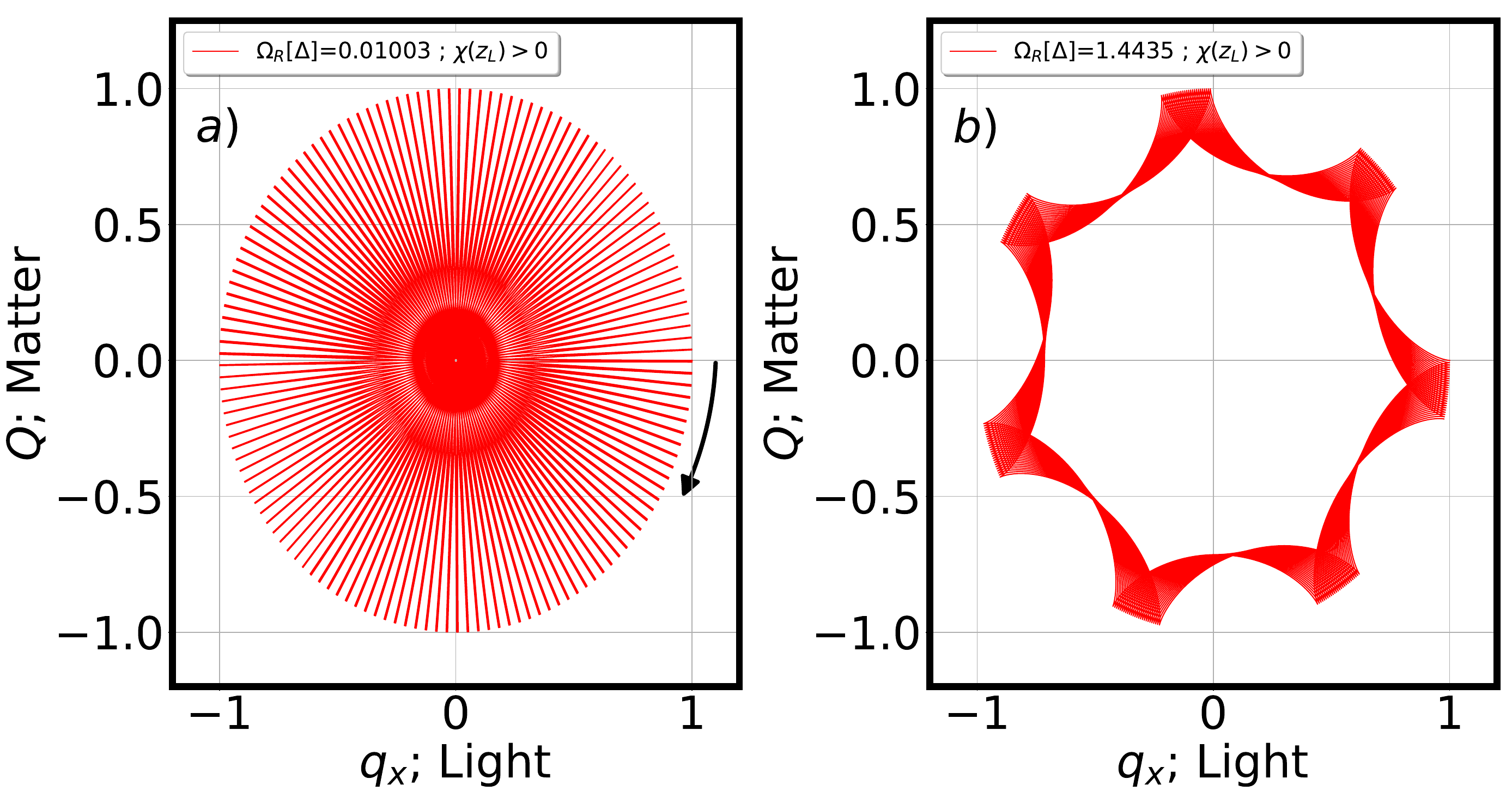}
\caption{
a) Plot of the classical trajectory $\left(q_x(t),q_y(t),Q(t)\right)$ in the plane $\left(q_x,Q\right)$, that is solution of Newton Eq.~\ref{Newton_1}, during the time-interval $t \in \left\lbrack 0, 1280/\Delta \right\rbrack$.
The collective vacuum Rabi splitting is $\Omega_R\approx 0.01$, and the gyrotropic coupling $\chi(z_L)=\mathcal{R}=10^{-3}>0$.
b) Same plot, but with $\Omega_R\approx 1.44$ and during time $t \in \left\lbrack 0, 700/\Delta \right\rbrack$.
Parameters are: $\Delta=1.0$, $\omega_c=1.0$, $m=1$, $\alpha_u=0.0$, $\alpha=\frac{\pi}{2}$, $z_L=\frac{L}{4}$, $q_x(0)=1.0$, $q_y(0)=Q(0)=0.0$, and $\dot{q}_x(0)=\dot{q}_y(0)=\dot{Q}(0)=0.0$. 
\label{fig:Fig3}}
\end{figure}
As shown in Fig.~\ref{fig:Fig2}-b), the $\delta^{\rm{2D}}_\eta$-signals acquire a non-monotoneous behavior upon sweeping $\omega_c$, reaching a maximum of $\left\lvert\delta^{\rm{2D}}_\eta\right\rvert$ close to resonance ($\omega_c \approx \Delta$).
For completeness, we present in Figs.~\ref{fig:Fig2}-a), and b), the computed $\delta^{\rm{3D}}_\eta$-signals for the 3D-bulk configuration (shown as inset in Fig.~\ref{fig:Fig2}-b)) multiplied by $10^2$ (magenta and cyan triangles and crosses) as a function of both $\Omega^{\rm{res}}_R = \Omega_R/\sqrt{2}$ and $\omega_c$. 
Surprisingly, in contrast to the $\rm{2D}$-case, the $\delta^{\rm{3D}}_\eta$-signals are much weaker and do not change sign upon sign-reversal of $\mathcal{R}$.
This is due to the fact that $\delta^{\rm{3D}}_\eta$ is obtained by integration of opposite terms $\delta^{\rm{2D}} _\eta(z_L)$ and $\delta^{\rm{2D}}_\eta(L-z_L)=-\delta^{\rm{2D}}_\eta(z_L)$ with respect to the plane of symmetry $z_L=L/2$ of the chosen ($m=1$) cavity-mode.
This argument can be generalized for any mode $m \in \mathbb{N}^*$.
This leads finally to a cancellation of all odd-contributions in $\mathcal{R}$ in the expression of $\delta^{\rm{3D}}_\eta \propto \mathcal{R}^2$, and to the relation $\delta^{\rm{3D}}_\eta(\mathcal{R})=\delta^{\rm{3D}}_\eta(-\mathcal{R})$ (see \cite{SupMat}).
We note that this behavior is consistent with the discussion following Eq.~\ref{Gyrotropic_Coupling}, and with Refs.~\cite{mauro_chiral_2023,mauro_classical_2024} that have shown in the framework of classical Maxwell equations, the absence of cavity-amplified chiroptical signals emerging at normal incidence of a 3D-bulk cavity made of normal metallic mirrors.
The crucial role played by dimensionality in the antisymmetry properties of the computed $\delta _\eta$-signals is one of the main results of this Letter.
%

%
%
\textit{Classical analogy.---} In order to further reveal their scope and interpret the previous results, we investigate the classical limit of the microscopic Hamiltonian modelling the FP cavity.
We show in \cite{SupMat} that for the 2D-layer (3D-bulk) case, this limit describes the classical motion of a point particle in dimension 2+1 (3+1), with unit mass and charge.
In the rest of the paper, we focus on the 2D-layer configuration, for which the effect of reduced dimensionality on the $\delta^{\rm{2D}}_\eta$-signals was revealed above, and the effective classical dynamics in dimension 2+1 can be easily computed and interpreted in phase-space.  
The coordinate vector of the fictive particle is in this case $\vec{x}(t) = q_x(t) \vec{e}_x + q_y(t) \vec{e}_y + Q(t) \vec{e}_z$, with $(q_x(t),q_y(t))$ the coordinates associated to the TE-modes and $Q(t)$ the one related to collective matter-excitation.
The fictive particle moves according to the Newton equations \cite{SupMat}
\begin{eqnarray}
\ddot{\vec{x}}(t) &=& - \underline{\underline{k}}_{\mathcal{R}}\vec{x}(t) + \dot{\vec{x}}(t) \times \vec{B}_{\mathcal{R}}
\label{Newton_1} \, .
\end{eqnarray}
with $\underline{\underline{k}}_{\mathcal{R}}=\mbox{ diag}\left(\omega^2_c,\omega^2_c,\Delta^2\right)+\delta\underline{\underline{k}}_{\mathcal{R}}$ a spring-tensor corresponding to an anisotropic elastic restoring-force, and 
\begin{eqnarray}
\delta\underline{\underline{k}}_{\mathcal{R}} &=&
4 \frac{\omega_c}{\Delta} \mathcal{R}^2 \Omega^2_R \cos^2\left(\theta_{z_L} \right)
\left[\begin{array}{ccc}
- v^2_y  & v_x v_y & 0 \\
v_x v_y & - v^2_x & 0 \\
0 & 0 & 0
\end{array}\right]
\label{Newton_2} \, .
\end{eqnarray}
The particle is also subject to an effective Lorentz-force, with magnetic field
\begin{eqnarray}
\vec{B}_{\mathcal{R}} &=& - 2 \sqrt{\frac{\Delta}{\omega_c}}\vec{e}_z \times \vec{\Omega}_{\rm{P},\mathcal{R}}
\, , 
\label{Newton_3} 
\end{eqnarray}
with $\vec{\Omega}_{\rm{P},\mathcal{R}} = \Omega_R \left\lbrace \sin\left(\theta_{z_L} \right) \hat{u}_\perp - \mathcal{R} \frac{\omega_c}{\Delta}\cos\left(\theta_{z_L} \right) \vec{e}_z \times  \hat{v}_\perp \right\rbrace$, the precession vector.
In absence of gyrotopic coupling ($\mathcal{R}=0$) the spring-tensor $\underline{\underline{k}}_0$ is diagonal and encodes free oscillations of the $q_{\alpha=x,y}(t)$ and $Q(t)$ modes at respective frequencies $\omega_c$ and $\Delta$.
The $\vec{B}_0$-field $\propto \Omega_R \vec{e}_z \times \hat{u}_\perp$ stands for a Larmor-like precession due to the the molecules-cavity electric-dipole interactions that dominate upon entering the achiral USC regime. 
At finite values of $\mathcal{R}$, magnetic-dipole interactions induce a correction to the $\vec{B}_\mathcal{R}$-field $\propto \mathcal{R} \Omega_R \hat{v}_\perp$, at the origin of gyrotropic coupling. 
The $\delta\underline{\underline{k}}_{\mathcal{R}}$-tensor is of lower-order $\propto \mathcal{R}^2$, and encodes for a negligible renormalization of $\omega_c$ and elastic-coupling between the cavity-modes. 
The Newton Eqs.~\ref{Newton_1} are integrable. 
Their solution is a linear combination of three eigenmodes of frequencies coinciding with the Hopfield polaritonic spectra (see \cite{SupMat}), thus confirming the essential classical nature of the polaritons \cite{zhu_vacuum_1990,hopfield_theory_1958}.
We also note that Eq.~\ref{Newton_1} shows similarities (although being not the same) with the dynamics of an harmonically trapped and rotating Bose gaz \cite{PhysRevLett.88.250403}. 
%

%
%
\textit{Analysis of classical trajectories.---} 
%
%
\begin{figure}[tb]
\includegraphics[width=1.0\linewidth, angle=-0]{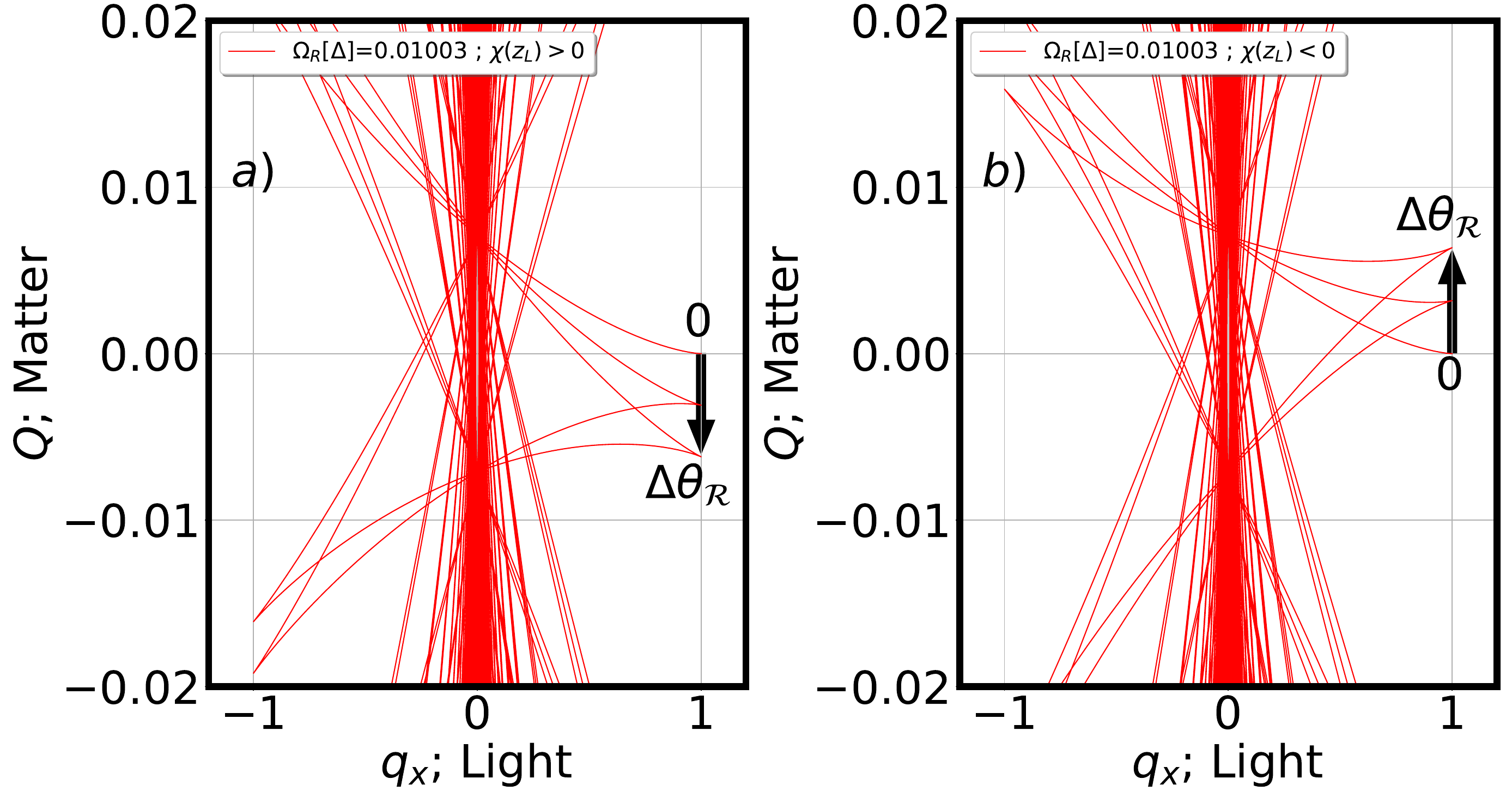}
\caption{
Same plots as in Fig.~\ref{fig:Fig3}, but with $\Omega_R\approx 0.01$.
a) Case with $\mathcal{R}=10^{-3}>0$. 
The black arrow shows the anholonomy angle $\Delta\theta_{\mathcal{R}}\approx \Delta\Delta\theta_{\mathcal{R}}>0$ given by Eq.~\ref{Hannay} for this range of parameters.
b) Case of opposite $\mathcal{R}=-10^{-3}<0$. The angle $\Delta\theta_{\mathcal{R}}<0$ reverses sign.
\label{fig:Fig4}}
\end{figure}
For arbitrary initial conditions, the obtained classical trajectories can be quite complex: they are indeed quasi-periodic and bounded (for non-vanishing restoring-force) but not closed \cite{arnol2013mathematical}, unless some specific commensurate values of the eigenmode frequencies are chosen.
In the following we integrate numerically Eqs.\ref{Newton_1} with initial conditions $q_x(0)=1.0$, $q_y(0)=Q(0)=0.0$, $\dot{q}_x(0)=\dot{q}_y(0)=\dot{Q}(0)=0.0$, and a choice of parameter $\alpha_u=0.0$ which enable an easy analysis of the trajectories (see \cite{SupMat} for another choice).
The obtained trajectories are shown in Fig.~\ref{fig:Fig3}, for $\omega_c=\Delta$ and $\chi_({z_L})>0$.
The point $\vec{x}(t)$ stays in the plane $(q_x,Q)$ and undergoes beat oscillations, with two characteristic frequencies: $\tilde{\omega}_{c,\mathcal{R}} = \sqrt{\omega^2_c + \Omega^2_{\rm{P},\mathcal{R}}}$ the renormalized cavity frequency, and $\Omega_{\rm{P},\mathcal{R}} = \left\lVert \vec{\Omega}_{\rm{P},\mathcal{R}} \right\rVert$ the precession frequency, depending on both the cVRS and gyrotropic coupling.
In the case of low cVRS $(\Omega_R \ll \omega_c)$, the resulting trajectory is equivalent to the motion of a fictive Foucault pendulum located at North pole (see \cite{SupMat}), with fast oscillation frequency $\tilde{\omega}_{c,\mathcal{R}} \approx \omega_c$.
We show in Fig.~\ref{fig:Fig3}-a) the effective pendulum trajectory, the plane of which rotates clockwise (black arrow) from the initial point $\vec{x}(0)=(1,0,0)$ with a slow precession frequency $\Omega_{\rm{P},\mathcal{R}}$. 
Waiting for times larger than the precession time ($t \gg 2\pi/\Omega_{\rm{P},\mathcal{R}}$), the trajectory of the classical mobile becomes dense in the plane $(q_x,Q)$ due to its already noticed quasi-periodic nature. 
We complement our analysis by considering the case of large cVRS ($\Omega_R > \omega_c$), when the system enters the USC.
We show in Fig.~\ref{fig:Fig3}-b), that the trajectories are dramatically modified in this regime.
Indeed, the fictive particle is now following an hypocycloid trajectory \footnote{See for instance \url{https://mathcurve.com/courbes2d/hypocycloid/hypocycloid.shtml}}, which converges to fast cyclotron gyrations around the direction of the effective $\vec{B}_{\mathcal{R}}$-field at very large $\Omega_R/\omega_c$ (see \cite{SupMat}).
At long times, the trajectory gets dense inside a ring-manifold, the lower circle of which has a radius $R_{\rm{min}} = \Omega_{\rm{P},\mathcal{R}}/\tilde{\omega}_{c,\mathcal{R}}$, while the upper circle radius is fixed by initial conditions and conservation of mechanical energy. 
The hole at the center of this ring-manifold is thus a striking topological signature of entering the USC regime. This USC topology turns out to be endowed with chiral features that we now describe.
%

%
\textit{Anholonomy of the classical trajectories.---} 
In order to unveil the role of gyrotropic coupling on the previous trajectories, we provide in Fig.~\ref{fig:Fig4} a zoom of Fig.~\ref{fig:Fig3}-a) close to the region $Q = 0$.
The initial point $\vec{x}(0)$ is labelled $0$, and the black arrow visualizes the change in angle $\Delta\theta_{\mathcal{R}}$ of the inertial plane of oscillations, after one full precession of the fictive Foucault pendulum.
The latter Hannay's angle \cite{hannay_angle_1985}, is a signature of anholonomy of the classical trajectories, and has some deep topological meaning \cite{berry_geometric_1988,berry1990quantum, anandan1992geometric,delplace_geometry_2020}.
We define the relative change in Hannay's angle $\Delta\Delta\theta_{\mathcal{R}}=\Delta\theta_{\mathcal{R}}-\Delta\theta_{0}$, as a quantitative measure of the gyrotropic coupling contribution to $\Delta\theta_{\mathcal{R}}$, that follows from a bit faster (slower) precession frequency due to positive (negative) $\chi(z_L)$.
We choose parameters such that $\left\lvert\Delta\theta_{0}\right\rvert \ll \left\lvert\Delta\Delta\theta_{\mathcal{R}}\right\rvert$, and visualize directly the quantity $\Delta\Delta\theta_{\mathcal{R}}\approx\Delta\theta_{\mathcal{R}}$ for which we have an analytical expression (see \cite{SupMat}) 
\begin{eqnarray}
\Delta\Delta\theta_{\mathcal{R}} &=& 2\pi\mathcal{R} \mbox{ cotan}\left( \theta_{z_L} \right)
\frac{\vec{e}_z \cdot \left( \vec{u} \times \vec{v}\right)}{\left\lVert \vec{u}_\perp \right\rVert^2}
\label{Hannay} \, .
\end{eqnarray}
Using Eq.~\ref{Gyrotropic_Coupling}, we see that $\Delta\Delta\theta_{\mathcal{R}}$ is directly proportional to the gyrotropic coupling parameter $\chi(z_L)$, namely that $\Delta\Delta\theta_{\mathcal{R}} = \pi \chi(z_L)/\left\lVert \vec{u}_\perp \right\rVert^2 \sin^2\left( \theta_{z_L} \right)$.
This relation is the major result of this Letter.
It connects the differential shift in the polaritonic spectra induced by gyrotropic coupling, to the variation of anholonomy angle of the classical trajectories in the analogous classical model. 
We checked in Fig.~\ref{fig:Fig4}-a) that, $\Delta\Delta\theta_{\mathcal{R}}>0$ when $\chi(z_L)>0$, while in Fig.~\ref{fig:Fig4}-b), $\Delta\Delta\theta_{\mathcal{R}}<0$ when $\chi(z_L)<0$.
%

%
In conclusion, we have shown the crucial role played by dimensionality in the mechanism of chiral light-matter interactions developing inside an optical Fabry-P\'erot cavity made of standard, achiral, mirrors.
We predicted that when the coupled molecules are arranged in a mesoscopic 2D-chiral layer, the emergence of a gyrotropic coupling is at the origin of a differential shift of the polaritonic spectra that are enantiodependent.
The latter effect results from a subtle interplay between the orientations, the enantiomeric class of the molecules, and the spatial dispersion of the cavity-modes along the optical axis. 
For macroscopic 3D-chiral bulk samples, such differential shift disappears. 
We interpreted this 2D-chiral effect by developing an analogous classical Newtonian model in dimension 2+1. 
We have shown that the molecular gyrotropic coupling is directly responsible for a perturbation of the anholonomy angle of the analogous classical trajectories.
The predicted $\delta^{\rm{2D}}_\eta$-signals while very weak, could in principle be observed with state-of-the-art optical spectroscopy \cite{gautier_planar_2022}.
We anticipate the latter signal to be robust to partial orientational disorder, but to be destroyed upon full rotational averaging of the molecular orientations.
We hope that our work will stimulate new theoretical efforts to find, and analyze the nature and topology of analogous classical models of chiral light-matter interactions like the present one. 
Such efforts would thereby lead to revisiting our current interpretations of materials chiroptical properties in the ultrastrong coupling regime.
%

\section*{Acknowledgments}
\label{Acknowledgments}
R. Avriller thanks Cl\'ement Dutreix for useful discussions about Hannay's angle in the Foucault pendulum problem.
R. Avriller acknowledges financial support of the CNRS-Chimie Projet Itin\'erance @INC2023, \textit{“Etude de la Dynamique Stochastique de R\'eactions Chimiques en Cavit\'e Electromagn\'etique“}. 
This work is also part of the Interdisciplinary Thematic Institute QMat of the University of Strasbourg, CNRS, and Inserm, supported by the following programs: IdEx Unistra (Grant No. ANR-10-IDEX-0002), SFRI STRATUS Project (Project No. ANR-20-SFRI-0012), and USIAS (Grant No. ANR-10-IDEX- 0002-02), under the framework of the French Investments for the Future Program. Further support from the PEPR LUMA Tornado project, and from the IdEx of the University of Bordeaux / Grand Research Program GPR LIGHT are acknowledged.
%

%
\bibliography{biblio}
%

\end{document}


\title{Supplementary material to the paper: "Chirality and dimensionality in the ultrastrong light-matter coupling regime"}

%
\author{R.~Avriller}
\affiliation{Univ.~Bordeaux, CNRS, LOMA, UMR 5798, F-33405 Talence, France}
\email{remi.avriller@u-bordeaux.fr}
\author{C.~Genet}
\affiliation{University of Strasbourg and CNRS, CESQ and ISIS, UMR 7006, F-67000 Strasbourg, France}
%

\begin{abstract}
We provide additional information about the microscopic Hamiltonian describing the gyrotropic coupling of molecules embedded inside an optical Fabry-P\'erot cavity.
We solve numerically and analytically when possible, the resulting spectra for Hopfield polaritons, in both cases of 2D-layer and 3D-bulk configuration of coupled molecules inside the cavity.
We show how to perform the classical limit of such quantum models in dimension 2 (3), obtaining an analogous classical Newtonian model in dimension 2+1 (3+1) with the same eigenmode spectrum.
Finally, in the 2D-layer configuration, we analyze the classical trajectories in the analogous model.
We obtain hypocycloid trajectories, equivalent to the motion of a Foucault pendulum in the weak light-matter coupling regime, while exhibiting cyclotron gyrations upon entering the ultrastrong coupling regime.
We relate explicitly the gyrotropic coupling to the anholonomy angle of the obtained classical trajectories. 
\end{abstract}

\maketitle

\tableofcontents

%
\section{2D-layer configuration of molecules in the optical cavity}
\label{2DChiralityorderedHopfield}
In this section, we derive the polariton spectra for the 2D-layer configuration of coupled molecules confined inside the optical Fabry-P\'erot cavity.
\subsection{Hopfield polaritons}
\label{HopfieldPolChiral}
%
\begin{figure}[!htb]
\centering
\includegraphics[width=1.0\linewidth]{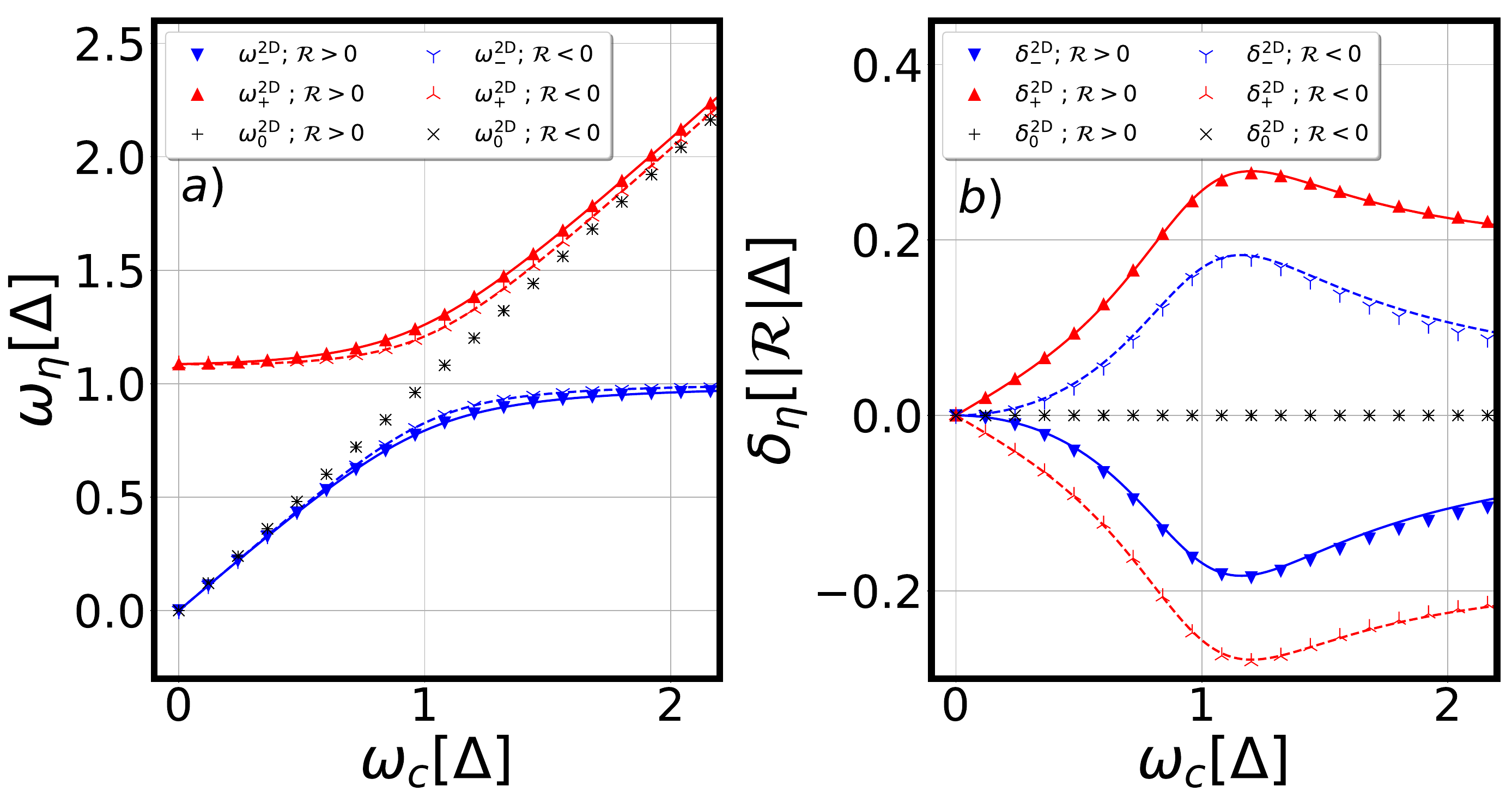}
\caption{
a) Spectra of the 2D-layer Hopfield-matrix in Eq.~\ref{2DHop8}, as a function of cavity frequency $\omega_c$.
Triangles and crosses are output of exact numerical diagonalisation, while plain or dashed-curves are the output of the analytical formula in Eq.~\ref{ChiralClassicalHbos2Modes20}. 
b) Same curves but for the relative shift in the polaritonic spectra $\delta^{\rm{2D}}_\eta = \omega^{\rm{2D}}_{\eta;\mathcal{R}} - \omega^{\rm{2D}}_{\eta;\mathcal{R}=0}$ (in units of $\left\lvert \mathcal{R} \right\rvert\Delta$).
Parameters are: $\Delta=1.0$, $m=1$, $z_L=\frac{L}{4}$, $\Omega_R=0.3$, $\alpha_u=\frac{\pi}{4}$, $\alpha=\frac{\pi}{2}$, and $\chi({z_L})=\mathcal{R}=0.1$.
}
\label{fig:FigS1}
\end{figure}
We first write explicitly the microscopic Hamiltonian describing the coupling of a 2D-layer of molecules to the cavity modes of the optical cavity.
The simplest theoretical model describing gyrotropic coupling, necessitates at least coupling the molecules to two cavity modes. 
We have chosen here for simplicity two $\mbox{TE}_{\vec{k}_\perp=0,m,x}$, and $\mbox{TE}_{\vec{k}_\perp=0,m,y}$ modes of degenerate frequency $\omega_c$ and different linear polarization $x$ and $y$ of the transverse electric field. 
More realistic and complex modelling taking into account all the optical modes (or a finite number of them larger than two) with finite $\vec{k}_\perp$-vectors is possible, but lay beyond the scope of the present paper, for which our model can be considered as a minimal one (but certainly not a complete one).
The molecules have an electric-dipole optical transition frequency $\Delta$, and couple through both their electric-transition and magnetic-transition dipole to the cavity-modes.
Note that, in order to investigate the regime of ultrastrong light-matter coupling, we have included into the Hamiltonian both counter-rotating terms and self-dipole interaction terms \cite{PhysRevB.72.115303}. 
Supposing that all the molecules couple in the same way to the TE-modes (they are supposed to have the same transition dipoles and molecular orientation, thus realizing an homogeneous aggregate), we write the following homogeneous microscopic Hamiltonian describing this problem 
(notations are those of the paper, for quantities already introduced in its main text)
\begin{eqnarray}
H &=& H_c + H_e + H_{D} + H_{M} + H_{SD}
\label{ChiralHcompact2Modes1} \, , 
\end{eqnarray}
with $H_c$ the Hamiltonian of the free cavity-modes
\begin{eqnarray}
H_c &=& \hbar\omega_c \left\lbrace 
a_{x}^\dagger a_{x} + a_{y}^\dagger a_{y} + 1 
\right\rbrace
\label{ChiralHcompact2Modes2} \, , 
\end{eqnarray}
with $a_{x(y)}$ ($a^\dagger_{x(y)}$), the bosonic operator that destroys (creates) an excitation in the mode $\mbox{TE}_{\vec{k}_\perp=0,m,x(y)}$.
The Hamiltonian of the free molecules is written 
\begin{eqnarray}
H_e &=& \hbar\frac{\Delta}{2} \sum_{j=1}^N \left\lbrace 
c_{e_j}^\dagger c_{e_j} - c_{g_j}^\dagger c_{g_j} \right \rbrace
\label{ChiralHcompact2Modes3} \, , 
\end{eqnarray}
The fermionic operators $c^\dagger_{g_j}$ and $c^\dagger_{e_j}$ create respectively a fermion on the ground-state $\ket{g_j}$ and excited-state $\ket{e_j}$ of molecule $j$.
The molecules interact with the cavity-modes through their electric-transition dipole interaction Hamiltonian $H_D$, and their magnetic-transition dipole Hamiltonian $H_M$
\begin{eqnarray}
H_D &=& -i\hbar \sum_{j=1}^N \left( b_j + b^\dagger_j \right)
\left\lbrace
g_{jx} \left( a_{x} - a_{x}^\dagger \right)
+ g_{jy}
\left( a_{y} - a_{y}^\dagger \right)
\right\rbrace 
\nonumber \, , \\
\label{ChiralHcompact2Modes4} \\
H_M &=& -i\hbar \sum_{j=1}^N \left( b_j - b^\dagger_j \right)
\left\lbrace
\nu_{jx} 
\left( a_{y} + a_{y}^\dagger \right)
- \nu_{jy} 
\left( a_{x} + a_{x}^\dagger \right)
\right\rbrace 
\nonumber \, , \\
\label{ChiralHcompact2Modes5}  
\end{eqnarray}
where we introduced $b^\dagger_{j}=c^\dagger_{e_j}c_{g_j}$ the operator that creates a (composite) exciton on molecule number $j$, and
\begin{eqnarray}
\hbar \vec{g}_j &\equiv & \sqrt{\frac{\hbar\omega_c}{\varepsilon_0 V}}\mbox{ }\vec{\mu}_{j} \sin\left( \theta_{z_j} \right) \in \mathbb{R}^3
\label{ChiralHcompact7} \, , \\
\hbar \vec{\nu}_j &\equiv & - i \frac{1}{c} \sqrt{\frac{\hbar\omega_c}{\varepsilon_0 V}} \mbox{ } \vec{m}_{j}\cos\left( \theta_{z_j} \right)
\in \mathbb{R}^3
\label{ChiralHcompact8} \, .
\end{eqnarray}
respectively the electric-dipole and magnetic-dipole coupling vectors, with $\theta_{z_j} = m \frac{\pi}{L}z_j$.
We consider now the case of homogeneous couplings $\vec{g}=g \vec{u}$ and $\vec{\nu}= \mathcal{R} g \vec{v}$.
The results written in the paper for the 2D-layer, can be obtained from the homogeneous case studied in this supplementary material, by multiplying the expressions involving $\vec{g}$ and $\vec{\nu}$ by $\sin\left( \theta_{z_j} \right)$ and $\cos\left( \theta_{z_j} \right)$ respectively (see Eqs.~\ref{ChiralHcompact7},\ref{ChiralHcompact8}), that depend on the spatial dispersion of the cavity-modes along the optical axis. 
Finally, we take into account self-dipole interactions $H_{SD}$ that are necessary to deal with the ultrastrong coupling regime
\begin{eqnarray}
H_{SD} &=& \frac{\hbar}{\omega_c} \left\lVert \hat{g}_\perp \right\rVert^2 
\left\lvert \sum^N_{j=1} \left( b_j + b^\dagger_j \right) \right\rvert^2 
\label{ChiralHcompact2Modes6} \, .
\end{eqnarray}
It is natural to further introduce the collective exciton (composite-boson) mode
\begin{eqnarray}
B = \frac{1}{\sqrt{N}}\sum_{j=1}^N b_j
\label{2DHop1} \, . 
\end{eqnarray}
In the limit of large $N\gg 1$, when the density of electronic excitation is very small $\sum_{i=1}^N \left\langle c_{ei}^\dagger c_{ei} \right\rangle \ll N$, the $B$-operator is approximately bosonic
\begin{eqnarray}
\left\lbrack B, B^\dagger \right\rbrack &\approx& 1
\label{2DHopbis} \, . 
\end{eqnarray}
Consistently, we bozonize the previous Hamiltonian, which results in a simpler quadratic Hamiltonian $H_{\rm{bos}}$ (Hopfield-Hamiltonian \cite{PhysRev.112.1555})
\begin{eqnarray}
H_{\rm{bos}} &=& H_c + H_e + H_{D} + H_{M} + H_{SD}
\label{ChiralHbos2Modes1} \, , \\
H_c &=& \hbar\omega_c \left\lbrace 
a_{x}^\dagger a_{x} + a_{y}^\dagger a_{y} + 1 
\right\rbrace
\label{ChiralHbos2Modes2} \, , \\
H_e &=& \hbar \Delta \sum_{j=1}^N \left\lbrace b_{j}^\dagger b_{j} + \frac{1}{2} \right\rbrace
\label{ChiralHbos2Modes3} \, , \\
H_D &=& -i\hbar \Omega_R \left( B + B^\dagger \right)
\left\lbrace
u_{x} \left( a_{x} - a_{x}^\dagger \right)
+ u_{y}
\left( a_{y} - a_{y}^\dagger \right)
\right\rbrace 
\nonumber \, , \\
\label{ChiralHbos2Modes4} \\
H_M &=& -i\hbar \mathcal{R} \Omega_R \left( B - B^\dagger \right) \left\lbrace
v_{x} 
\left( a_{y} + a_{y}^\dagger \right)
- v_{y} 
\left( a_{x} + a_{x}^\dagger \right)
\right\rbrace 
\nonumber \, , \\
\label{ChiralHbos2Modes5}  \\
H_{SD} &=& \hbar D
\left( B + B^\dagger \right)^2
\label{ChiralHbos2Modes6} \, ,
\end{eqnarray}
where we introduced
\begin{eqnarray}
\Omega_R &=& g \sqrt{N}
\label{2DHop2} \, , \\
D &=& \frac{\Omega_R^2}{\omega_c} \left\lVert \vec{u}_\perp \right\rVert^2 
\label{2DHop3} \, ,
\end{eqnarray}
respectively the collective vacuum Rabi splitting (cVRS) and self-dipole interaction terms.
The diagonalization of $H_{\rm{bos}}$ amounts to find eigenoperators $\chi$ (which are bosonic destruction operators) of the former quadratic Hamiltonian, in the form of the following linear combinations of destruction and creation operators (Bogoliubov transformation)\cite{PhysRev.112.1555}
\begin{eqnarray}
\chi &=& \underline{\hat{A}}^t \cdot \underline{C}
\label{2DHop4} \, , \\
\underline{\hat{A}} &=& \left\lbrack
a_x, a_y, B, a^\dagger_x, a^\dagger_y, B^\dagger
\right\rbrack
\label{2DHop5} \, , \\
\underline{C} &=& \left\lbrack
\alpha_x, \alpha_y, \gamma, \beta_x, \beta_y, \delta
\right\rbrack
\label{2DHop6} \, ,
\end{eqnarray}
with $\underline{C}$ the vector of coefficients of this linear combination. 
The $\chi$-operators follow a linear eigenvalue problem, 
\begin{eqnarray}
\omega \underline{\hat{A}}^t\cdot \underline{C} &=& \underline{\hat{A}}^t\cdot \underline{\underline{\mathcal{H}}} \cdot \underline{C}
\label{Bog1}  \, , \\
\underline{C}^{\dagger} \cdot \underline{\underline{g}} \cdot \underline{C} &=& 1 \label{Bog2}  \, , 
\end{eqnarray}
with the $\underline{\underline{g}}$-metric 
\begin{eqnarray}
\underline{\underline{g}} &=& \mbox{diag} \left( 1,1,1,-1,-1,-1 \right)
\label{2DHop7}  \, ,
\end{eqnarray}
and Hopfield-matrix $\underline{\underline{\mathcal{H}}}$ of dimension $6\times 6$
%
\begin{widetext}
\begin{eqnarray}
\underline{\underline{\mathcal{H}}} &=&
\left[\begin{array}{cccccc}
\omega_c & 0 & -i\Omega_R(u_x+\mathcal{R}v_y) & 0 & 0 & i\Omega_R (u_x-\mathcal{R}v_y)	\\
0 & \omega_c & -i\Omega_R(u_y-\mathcal{R}v_x) & 0 & 0 & i\Omega_R (u_y + \mathcal{R}v_x) \\
i\Omega_R(u_x+\mathcal{R}v_y) & i\Omega_R(u_y-\mathcal{R}v_x) & \Delta + 2D & i\Omega_R(u_x-\mathcal{R}v_y) & i\Omega_R(u_y+\mathcal{R}v_x) & -2D \\
0 & 0 & i\Omega_R(u_x-\mathcal{R}v_y) & -\omega_c & 0 & -i\Omega_R (u_x+\mathcal{R}v_y) \\
0 & 0 & i\Omega_R(u_y+\mathcal{R}v_x) & 0 & -\omega_c & -i\Omega_R (u_y-\mathcal{R}v_x) \\
i\Omega_R(u_x-\mathcal{R}v_y) & i\Omega_R(u_y+\mathcal{R}v_x) & 2D & i\Omega_R(u_x+\mathcal{R}v_y) & i\Omega_R(u_y-\mathcal{R}v_x) & -\left( \Delta + 2D \right) 
\end{array}\right]
\nonumber \, . \\
\label{2DHop8}
\end{eqnarray}
\end{widetext}
%
The spectrum of eigenvalues of $\underline{\underline{\mathcal{H}}}$ is found by numerical diagonalization.
We find $3$ positive eigenvalues of the $\underline{\underline{\mathcal{H}}}$-matrix: one lower polariton $\omega^{\rm{2D}}_{-}$, one upper polariton $\omega^{\rm{2D}}_{+}$, and one middle polariton $\omega^{\rm{2D}}_{0}$.
We note that the characteristic polynomial $P(\lambda) = \mbox{det}\left( \lambda \underline{\underline{\mbox{Id}}} - \underline{\underline{\mathcal{H}}} \right)$ of the Hopfield-matrix can be obtained analytically as
\begin{eqnarray}
P(\lambda) &=& \left( \lambda^2-\omega_c^2 \right)^2\left( \lambda^2-\Delta^2 \right)
-
4  \frac{\Omega_R^2}{\omega_c}\left( \lambda^2-\omega_c^2 \right)
\Big\lbrace
\nonumber \\
&&\Delta \left\lVert \vec{u}_\perp \right\rVert^2 \lambda^2
+
2 \mathcal{R} \lambda^2 \omega_c \vec{e}_z\cdot\left( \vec{u}_\perp \times \vec{v}_\perp \right)
\nonumber \\
&+& \mathcal{R}^2 \Delta\omega_c^2 \left\lVert \vec{v}_\perp \right\rVert^2
\Big\rbrace
- 16 \mathcal{R}^2 \lambda^2 \Omega_R^4
\left( \vec{u}_\perp \cdot \vec{v}_\perp \right)^2
\nonumber \, . \\
\label{2DHop9}  
\end{eqnarray}
We show in Fig.~\ref{fig:FigS1} the polaritonic spectra $\omega^{\rm{2D}}_{\eta}$ and differential spectra $\delta^{\rm{2D}}_\eta = \omega^{\rm{2D}}_{\eta;\mathcal{R}} - \omega^{\rm{2D}}_{\eta;\mathcal{R}=0}$, at fixed $\Omega_R = 0.3 \Delta$, as a function of the mode frequency $\omega_c$.
Compared to Fig.2 of the paper, we chose a rather large value of $\mathcal{R}=0.1$, to emphasize the effect of gyrotropic coupling on the global $\omega^{\rm{2D}}_{\eta}$-signals.
We also added the case of the middle polariton (black crosses) $\omega^{\rm{2D}}_{0}$ which do not exhibit any differential shift at finite $\mathcal{R}$ ($\delta^{\rm{2D}}_0 = 0$).
Despite the large value of $\mathcal{R}=0.1$ chosen, we still get a very good agreement between the numerics (triangle and cross-curves) and the analytical formula (plain and dashed-curves) in Eq.~\ref{ChiralClassicalHbos2Modes20}.
Small differences between them are due to the range of validity of the analytics, which was assigned to be of order $o(\mathcal{R})$.
The same curves are shown in Fig.~\ref{fig:FigS2} at fixed $\omega_c=\tilde{\Delta}=\sqrt{\Delta^2 + 4\frac{\Delta}{\omega_c}\Omega^2_R \left\lVert \vec{u}_\perp \right\rVert^2}$, as a function of the collective Vaccum Rabi Splitting $\Omega^{\rm{res}}_{R}= \Omega_R$.
For completeness, we show in Fig.~\ref{fig:FigS3} the same plots as a function 
of the 2D-layer position $z_L$, but at fixed $\omega_c = \tilde{\Delta}$, $\Omega^{\rm{res}}_R=0.3 \Delta$ and $\mathcal{R}=0.05$.
We see the crucial role played by the spatial dispersion of the mode $m=1$ along the optical axis, and the relative position of the 2D-layer along that profile.
Interestingly, the differential spectra $\delta^{\rm{2D}}_\eta(z_L)$ are minimum at the center of the mode $z_L=L/2$ where they vanish ($\delta^{\rm{2D}}_\eta(L/2)=0$), and maximum in regions of maximum inhomogeneity ($z$-variation) of the electromagnetic cavity-fields.
The agreement between numerics and analytics is still very good, although some discrepancies are seen close to the edges of the mode $z_L\approx 0,L$ where the  corrections of higher-order in $\mathcal{R}$ cannot be neglected anymore in the analytics. 
\subsection{Classical limit of the Hopfield model}
\label{Classical_Model_dim_3}
%
\begin{figure}[!htb]
\centering
\includegraphics[width=1.0\linewidth]{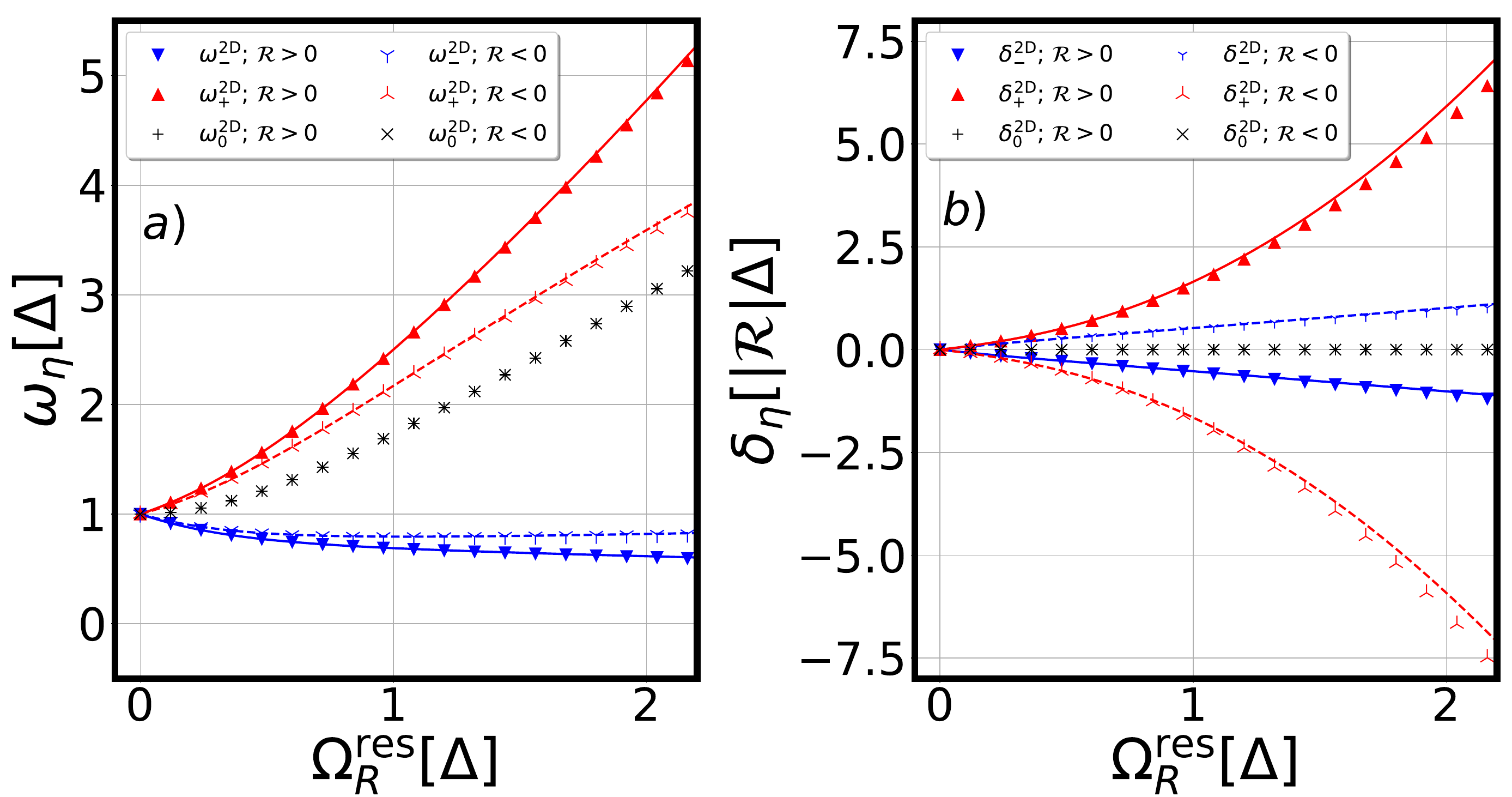}
\caption{
Same Fig. as Fig.~\ref{fig:FigS1}, but as a function of the collective Rabi splitting $\Omega^{\rm{res}}_{R} = \Omega_R$ at fixed $\omega_c=\tilde{\Delta}$. 
}
\label{fig:FigS2}
\end{figure}
\begin{figure}[!htb]
\centering
\includegraphics[width=1.0\linewidth]{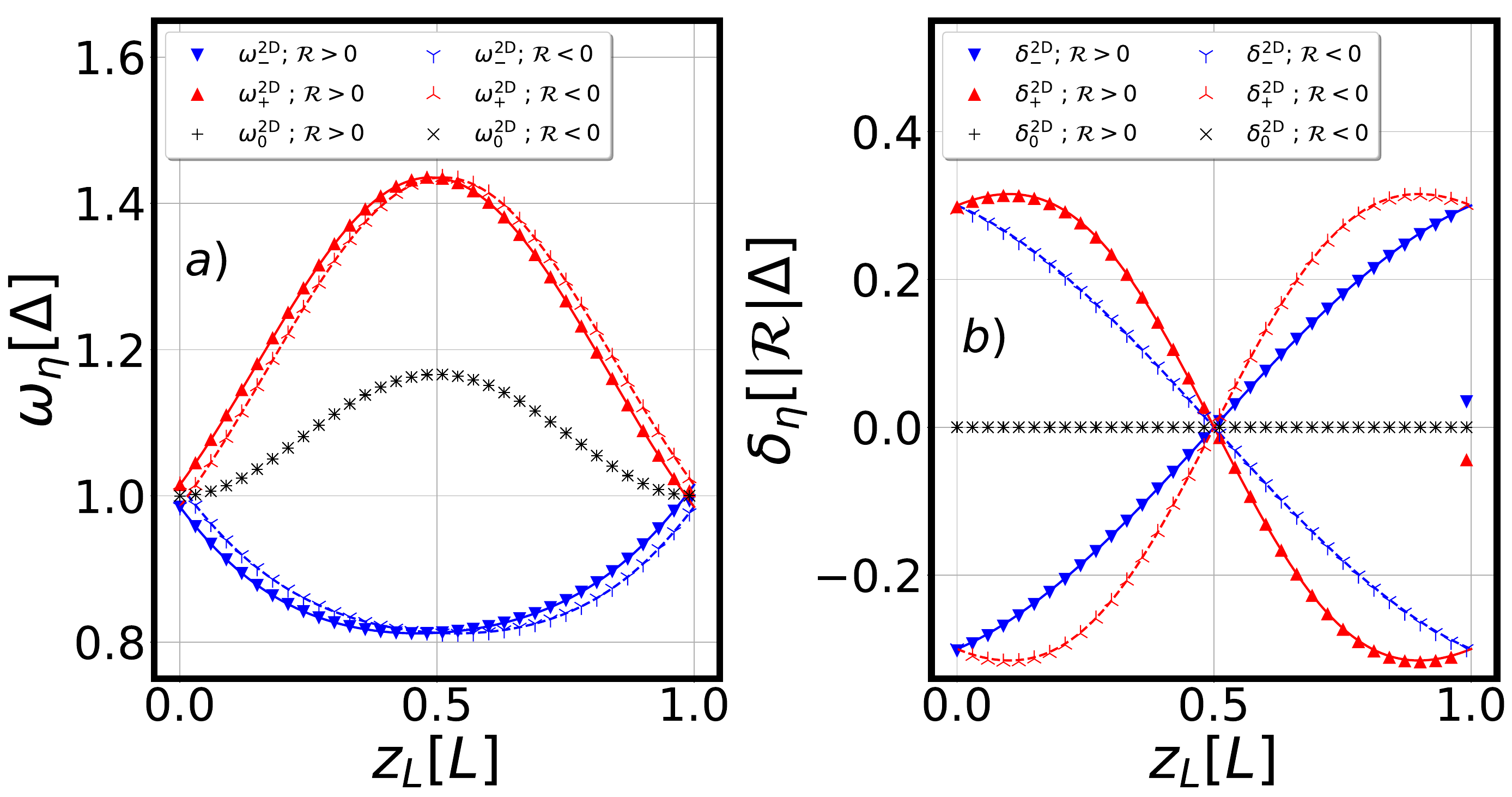}
\caption{
Same as Fig.~\ref{fig:FigS1}, but as a function of the 2D-layer position $z_L$, at fixed $\omega_c = \tilde{\Delta}$, $\Omega^{\rm{res}}_R=0.3 \Delta$ and $\mathcal{R}=0.05$. 
}
\label{fig:FigS3}
\end{figure}

The classical limit of the $H_{\rm{bos}}$-Hamiltonian in Eq.~\ref{ChiralHbos2Modes1}, is obtained as
\begin{eqnarray}
H^{(\rm{cl})}_{\rm{bos}} &=& \frac{\left\lVert \vec{p}_\perp \right\rVert^2}{2}
+ \frac{\omega^2_c}{2}\left\lVert \vec{q}_\perp \right\rVert^2
+ \frac{P^2}{2}
+ \frac{\Delta^2}{2} Q^2
\nonumber \\
&+&
\omega_P \vec{u}_\perp \cdot \vec{p}_\perp Q + 
\left( \vec{\Omega}_g \times \vec{v}_\perp \right) \cdot \vec{q}_\perp P
\nonumber \\
&+&
\frac{\omega^2_P}{2} \left\lVert \vec{u}_\perp \right\rVert^2 Q^2
\label{ChiralClassicalHbos2Modes1} \, , 
\end{eqnarray}
where we introduced the conjugated position and momentum variables
\begin{eqnarray}
q_{x(y)} &=& \sqrt{\frac{\hbar}{2\omega_c}} \left( a_{x(y)} + a^\dagger_{x(y)} \right)
\label{ChiralClassicalHbos2Modes2} \, , \\
p_{x(y)} &=& \sqrt{\frac{\hbar\omega_c}{2}} \frac{ a_{x(y)} - a^\dagger_{x(y)}}{i}
\label{ChiralClassicalHbos2Modes3} \, , \\
Q &=& \sqrt{\frac{\hbar}{2\Delta}} \left( B + B^\dagger \right)
\label{ChiralClassicalHbos2Modes4} \, , \\
P &=& \sqrt{\frac{\hbar\Delta}{2}} \frac{ B - B^\dagger}{i}
\label{ChiralClassicalHbos2Modes5} \, ,
\end{eqnarray}
as well as the quantities
\begin{eqnarray}
\omega_P &=& 2 \sqrt{\frac{N\Delta}{\omega_c}} g
\label{ChiralClassicalHbos2Modes6} \, , \\
\vec{\Omega}_g &=& \Omega_g \vec{e}_z = \mathcal{R}\frac{\omega_c}{\Delta} \omega_P \vec{e}_z 
\label{ChiralClassicalHbos2Modes7} \, . 
\end{eqnarray}
In the classical limit, the conjugate variables are commuting variables, that fulfil classical Hamilton equations
\begin{eqnarray}
\dot{\vec{q}}_\perp &=& \vec{p}_\perp + \omega_P Q \vec{u}_\perp
\label{ChiralClassicalHbos2Modes8} \, , \\
\dot{Q} &=& P + \left( \vec{\Omega}_g \times \vec{v}_\perp \right) \cdot \vec{q}_\perp
\label{ChiralClassicalHbos2Modes9} \, , \\
\dot{\vec{p}}_\perp &=& - \omega^2_c \vec{q}_\perp - \left( \vec{\Omega}_g \times \vec{v}_\perp \right) P 
\label{ChiralClassicalHbos2Modes10} \, , \\
\dot{P} &=& - \Delta^2 Q - \omega_P \vec{u}_\perp \cdot \vec{p}_\perp - \omega^2_P \left\lVert \vec{u}_\perp \right\rVert^2 Q
\label{ChiralClassicalHbos2Modes11} \, . 
\end{eqnarray}
From the latter equations, one can derive classical Newton equation of motions for the $(\vec{q}_\perp,Q)$ coordinates alone
\begin{eqnarray}
\ddot{Q} &=& - \Delta^2 Q - \left( \omega_P \vec{u}_\perp -\vec{\Omega}_g \times \vec{v}_\perp \right)\cdot \dot{\vec{q}}_\perp
\label{ChiralClassicalHbos2Modes13} \, , \\
\ddot{\vec{q}}_\perp &=& -\omega^2_c \vec{q}_\perp
+
\left(\vec{\Omega}_g \times \vec{v}_\perp \right)
\left\lbrack
\left(\vec{\Omega}_g \times \vec{v}_\perp \right)\cdot\vec{q}_\perp
\right\rbrack
\nonumber \\
&+&
\left( \omega_P \vec{u}_\perp -\vec{\Omega}_g \times \vec{v}_\perp \right)\dot{Q}
\label{ChiralClassicalHbos2Modes14} \, . 
\end{eqnarray}
We look for eigenmodes of the Newton equations~\ref{ChiralClassicalHbos2Modes13},\ref{ChiralClassicalHbos2Modes14}
\begin{eqnarray}
\vec{q}_\perp(t) &=& \vec{q}_\perp(\omega) e^{-i\omega t} 
\label{ChiralClassicalHbos2Modes15} \, , \\
Q(t) &=& Q(\omega) e^{-i\omega t} 
\label{ChiralClassicalHbos2Modes16} \, ,
\end{eqnarray}
thus leading to a system of 3 coupled linear equations
\begin{eqnarray}
&&\left( \omega^2 - \Delta^2 \right) Q(\omega) + i\omega \left( \omega_P \vec{u}_\perp -\vec{\Omega}_g \times \vec{v}_\perp \right)\cdot \vec{q}_\perp(\omega) = 0
\nonumber \\
\label{ChiralClassicalHbos2Modes17}\\
&&\left( \omega^2 -\omega^2_c \right) \vec{q}_\perp(\omega)
+
\left(\vec{\Omega}_g \times \vec{v}_\perp \right)
\left\lbrack
\left(\vec{\Omega}_g \times \vec{v}_\perp \right)\cdot\vec{q}_\perp(\omega)
\right\rbrack
\nonumber \\
&-&
i\omega\left( \omega_P \vec{u}_\perp -\vec{\Omega}_g \times \vec{v}_\perp \right) Q(\omega) = 0
\label{ChiralClassicalHbos2Modes18} \, . 
\end{eqnarray}
Non-trivial solutions of this system of equations are found as zeros of the secular-determinant $\Delta^{(\rm{cl})}(\omega)$
\begin{eqnarray}
\Delta^{(\rm{cl})}(\omega) &=& \left( \omega^2-\omega_c^2 \right)^2\left( \omega^2-\Delta^2 \right)
-
\frac{\omega_P^2}{\Delta}\left( \omega^2-\omega_c^2 \right)
\Big\lbrace
\nonumber \\
&&\Delta \left\lVert \vec{u}_\perp \right\rVert^2 \omega^2
+
2 \mathcal{R} \omega^2 \omega_c \vec{e}_z\cdot\left( \vec{u}_\perp \times \vec{v}_\perp \right)
\nonumber \\
&+& \mathcal{R}^2 \Delta\omega_c^2 \left\lVert \vec{v}_\perp \right\rVert^2
\Big\rbrace
- \mathcal{R}^2 \omega^2 \omega^2_c \frac{\omega_P^4}{\Delta^2}
\left( \vec{u}_\perp \cdot \vec{v}_\perp \right)^2
\nonumber \, . \\
\label{ChiralClassicalHbos2Modes19} 
\end{eqnarray}
Considering that $\omega^2_P = 4 \Delta \Omega^2_R/\omega_c$, it is simple to show that the determinant $\Delta^{(\rm{cl})}(\omega)$ in Eq.~\ref{ChiralClassicalHbos2Modes19} is the same as the characteristic polynomial $P(\omega)$ of the Hopfield-matrix written in Eq.~\ref{2DHop9}.
\textit{Thus both problems have the same eigensolutions and describe the same collective polariton spectra. 
It is remarkable that the solution of the quantum Bogoliubov problem in Sec.\ref{HopfieldPolChiral} is the same as the solution of the classical coupled-mode problem derived in this section.}
In the regime $\mathcal{R} \ll 1$, we provide a simple analytical solution for the polariton excitations $\eta=+(-)$ associated to the upper(lower) polariton, that are the zeros of Eq.~\ref{ChiralClassicalHbos2Modes19}
\begin{eqnarray}
\omega^{\rm{2D}}_\eta &\approx & \omega^{(0)}_\eta \left\lbrace
1 + \eta \mathcal{R}\left\lbrack
\vec{e}_z\cdot\left( \vec{u}_\perp \times \vec{v}_\perp \right)
\right\rbrack
\frac{\omega^2_P\omega_c}{2\Delta\tilde{\Omega}^2}
\right\rbrace
\label{ChiralClassicalHbos2Modes20} \, , \\
\omega^{(0)}_\eta &=&  \sqrt{
\frac{\omega^2_c + \tilde{\Delta}^2}{2} 
+ \eta \tilde{\Omega}^2
}
\label{ChiralClassicalHbos2Modes21} \, , \\
\tilde{\Omega}^2 &=& \sqrt{ 
\left( \frac{\omega^2_c - \tilde{\Omega}^2}{2} \right)^2 
+
\omega_c^2 \omega^2_P \left\lVert \vec{u}_\perp \right\rVert^2 } 
 \label{ChiralClassicalHbos2Modes22} \, , \\
\tilde{\Delta} &=& \sqrt{ \Delta^2 + \omega^2_P\left\lVert \vec{u}_\perp \right\rVert^2} 
\label{ChiralClassicalHbos2Modes23} \, . 
\end{eqnarray}
When $\mathcal{R} = 0$, we recover, as expected, the result of the Hamiltonian without gyrotropic coupling (see for instance Refs.~\cite{PhysRevB.72.115303,PhysRevLett.117.153601}), but with $\omega^2_P$ replaced by $\omega^2_P \left\lVert \vec{u}_\perp \right\rVert^2$: this is due to the contributions of both $\mbox{TE}_{x,y}$-modes that add-up quadratically.
For non-vanishing $\mathcal{R}$, Eq.~\ref{ChiralClassicalHbos2Modes20} provides the main analytical expression of the paper.
We show in Figs.~\ref{fig:FigS1},\ref{fig:FigS2},\ref{fig:FigS3}, that the outcome of the analytical formula written in Eq.~\ref{ChiralClassicalHbos2Modes20} matches very well with the outcome the numerically exact diagonalization of the Hopfield-matrix. 
\subsection{Analogous classical model in dimension $2+1$}
\label{Classical_Model_dim_3_Phy_Int}
%
%
We provide in this section a physical interpretation of the Newton equations~\ref{ChiralClassicalHbos2Modes13},\ref{ChiralClassicalHbos2Modes14} describing the dynamics of the cavity-modes $(q_x(t),q_y(t))$ coupled to the collective exciton-mode $Q(t)$.
For that purpose, we introduce the vector
\begin{eqnarray}
\vec{x}(t) &=& q_x(t) \vec{e}_x + q_y(t) \vec{e}_y 
+ Q(t) \vec{e}_z 
\label{Classical_Model_dim_3_Phy_Int1} \, ,
\end{eqnarray}
associated to \textit{the motion of a fictitious classical mobile of unit mass and unit charge. 
This mobile is subject to forces, and its dynamics is solution of the following Newton equation} (equivalent to Eqs.~\ref{ChiralClassicalHbos2Modes13},\ref{ChiralClassicalHbos2Modes14})
\begin{eqnarray}
\ddot{\vec{x}}(t) &=& - \underline{\underline{k}}\vec{x}(t) + \vec{M}\times \dot{\vec{x}}(t)
\label{Classical_Model_dim_3_Phy_Int2} \, ,
\end{eqnarray}
where we introduced the vector 
\begin{eqnarray}
\vec{M} &=& \Omega_g \vec{v}_\perp + \omega_p \vec{e}_z \times \vec{u}_\perp
\label{Classical_Model_dim_3_Phy_Int3} \, ,
\end{eqnarray}
and the anisotropic and symmetric spring-tensor
\begin{eqnarray}
\underline{\underline{k}} &=&
\left[\begin{array}{ccc}
\omega^2_c - \Omega^2_g v^2_y & \Omega^2_g v_x v_y & 0 \\
\Omega^2_g v_x v_y & \omega^2_c - \Omega^2_g v^2_x & 0 \\
0 & 0 & \Delta^2
\end{array}\right]
\label{Classical_Model_dim_3_Phy_Int4} \, ,
\end{eqnarray}
with $\Omega_g = \mathcal{R}\frac{\omega_c}{\Delta}\omega_P$.
The dynamics described by the Newton Eq.~\ref{Classical_Model_dim_3_Phy_Int2} is particularly simple and transparent. 
\textit{It describes the motion of a fictive classical mobile in dimension 2+1, that is subject to three types of forces}, namely
\begin{enumerate}
\item
\textit{Restoring spring-forces} that have different spring constants $(\omega^2_c - \Omega^2_g v^2_y,\omega^2_c - \Omega^2_g v^2_x,\Delta^2)$
along the different directions $(x,y,z)$, and are modified by gyrotropic coupling (of order $\mathcal{R}^2$). \\
\item
\textit{Elastic mode-coupling forces} $\propto\Omega^2_g v_x v_y$ coupling the $x$ and $y$ coordinates of the mobile, standing for an interaction between the $\mbox{TE}_x$ and $\mbox{TE}_y$ modes, mediated by the gyrotropic coupling (of order $\mathcal{R}^2$). \\
\item
\textit{A Lorentz-like force with effective magnetic B-field} $\vec{B} = -\vec{M}$.
This force is the sum of two contributions: the first one $\propto \omega_p \vec{e}_z \times \vec{u}_\perp$ is \textit{the Larmor-like coupling induced by electric-dipole coupling terms} in Eq.~\ref{ChiralHcompact2Modes4}.
The second contribution $\propto  \Omega_g \vec{v}_\perp$ is due to \textit{the magnetic-dipole interaction term} at the origin of the gyrotropic coupling (of order $\mathcal{R}$) in Eq.~\ref{ChiralHcompact2Modes5}.
\end{enumerate}
This analogy between the problem of gyrotropic coupling of a 2D-layer in a cavity with the dynamics of a classical mobile in dimension 2+1, is the main result of this section.
It is extremely simple and elegant.
\textit{In particular, the emergence of a Lorentz-force is a peculiarity of dimension 3, which arises because we considered here i) only 2 TE-modes coupling to the collective molecular exciton, and ii) a 2D-layer configuration of coupled molecules inside cavity.}
Including more optical modes in the description (for instance $M=4$ modes with 2 TE and 2 TM-modes) would result in a more general coupling to an anti-symmetric tensor with $M(M+1)/2$ (=$10$ for $M=4$) independent matrix elements, and would thus complexify the physical interpretation.  
\textit{The 2D-layer configuration is a simple case, where the Lorentz-force derived above has some effect on the topology in the classical trajectories describing the dynamics of the polaritons.
}
\subsection{Analysis of the classical trajectories}
\label{Foucault}
\subsubsection{Analytical solution for the trajectories}
\label{Exact}
%
%
%
\begin{figure}[tbh]
\includegraphics[width=1.0\linewidth, angle=-0]{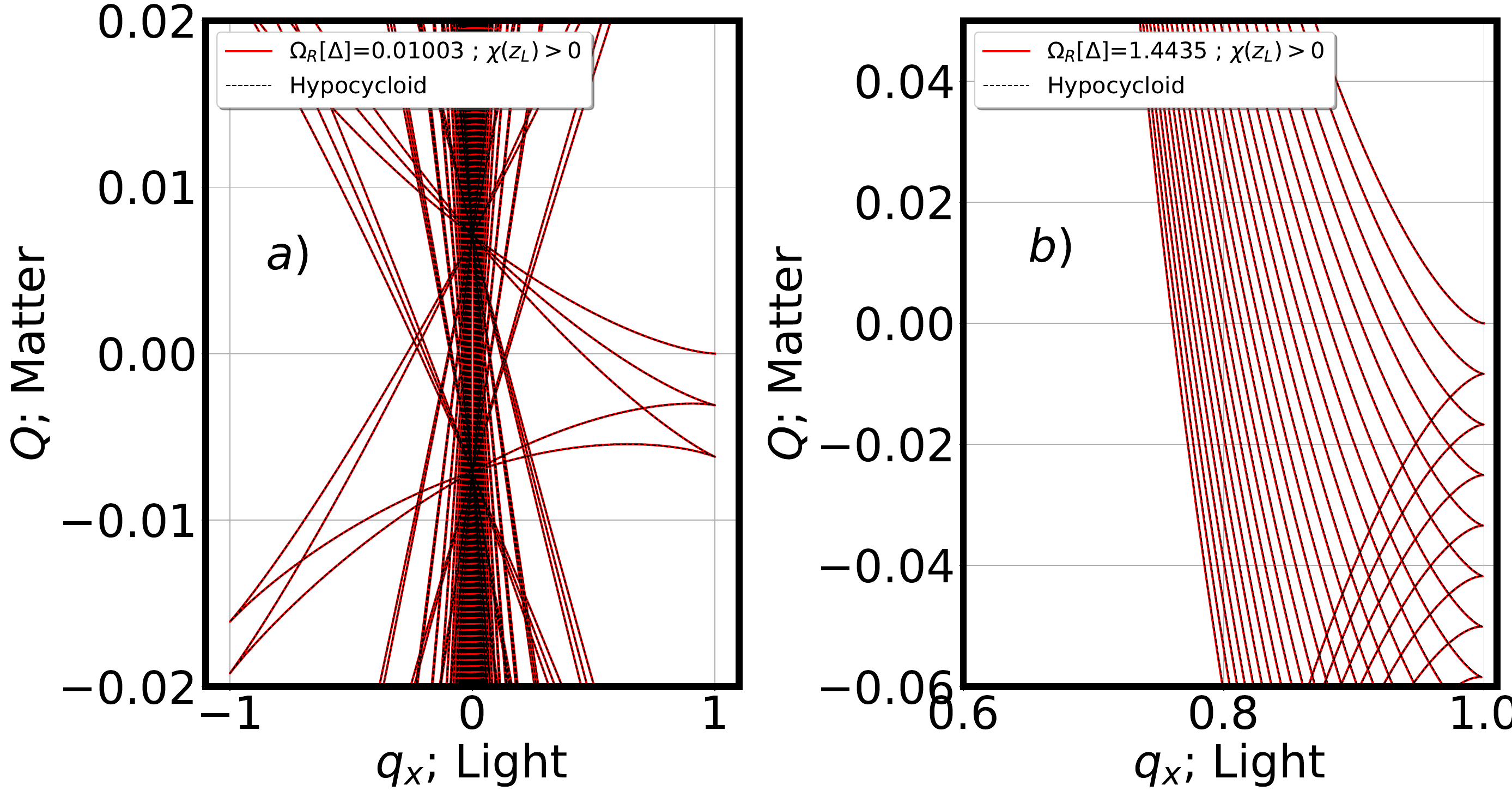}
\caption{
Plot of the classical trajectory $\left(q_x(t),q_y(t),Q(t)\right)$ in the plane $\left(q_x,Q\right)$ with the same parameters as in Fig.3 of the paper. 
Plain red curve: trajectory obtained by numerical integration of the Newton Eqs.~\ref{Classical_Model_dim_3_Phy_Int2}.
Dashed black curves: analytical solution from Eqs.~\ref{Strong_Coupling_1},\ref{Strong_Coupling_2},\ref{Strong_Coupling_3}, providing hypocycloid curves with parameters
given by Eqs.~\ref{Strong_Coupling_4},\ref{Strong_Coupling_5},\ref{Strong_Coupling_6}.
}
\label{fig:FigS7}
\end{figure}
We rewrite the Newton Eq.~\ref{Classical_Model_dim_3_Phy_Int2}, with a simpler notation $x(t)=q_x(t)$, $y(t)=q_y(t)$ and $z(t)=Q(t)$.
In the limit $\mathcal{R} \ll 1$ (realized in practise), we can neglect the elastic (non-diagonal) mode-coupling forces in the expression of the spring-tensor $\underline{\underline{k}}$ (see Eq.~\ref{Classical_Model_dim_3_Phy_Int4}), and obtain the following system of linear equations describing the motion of the fictive particle
\begin{eqnarray}
\ddot{x}(t) &\approx& - \omega^2_c x(t) + \Omega \cos\left( \beta \right) \dot{z}(t) 
\label{Foucault_1} \, , \\
\ddot{y}(t) &\approx& - \omega^2_c y(t) + \Omega \sin\left( \beta \right) \dot{z}(t) 
\label{Foucault_2} \, , \\
\ddot{z}(t) &\approx& - \Delta^2 z(t) - \Omega \left\lbrace
\cos\left( \beta \right) \dot{x}(t) + \sin\left( \beta \right) \dot{y}(t) 
\right\rbrace
\nonumber \, , \\
\label{Foucault_3} 
\end{eqnarray}
where we introduced the vector $\vec{\Omega} = \Omega \left\lbrace
\cos\left( \beta \right) \vec{e}_x + \sin\left( \beta \right)  \vec{e}_y
\right\rbrace$, with $\Omega = \left\lVert \vec{\Omega} \right\rVert^2$ and
\begin{eqnarray}
\vec{\Omega} &=& \vec{\Omega}_0 + \vec{\Omega}_1
\label{Foucault_4} \, , \\
\vec{\Omega}_0 &=& \omega_P \vec{u}_\perp
\label{Foucault_5} \, , \\
\vec{\Omega}_1 &=& - \vec{\Omega}_g \times \vec{v}_\perp
\label{Foucault_6} \, .
\end{eqnarray}
We wrote $\vec{\Omega}_0$ the contribution of the electric transition-dipole to $\vec{\Omega}$, and $\vec{\Omega}_1$ the one due to the magnetic transition-dipole.
The linear system of Eqs.~\ref{Foucault_1},\ref{Foucault_2},\ref{Foucault_3} can be simplified by doing a rotation $R_z\left( \beta \right)$ of angle $\beta$, in the plane $(x,y)$
\begin{eqnarray}
\left[\begin{array}{c}
X(t) \\
Y(t)
\end{array}\right] &=&  R_z\left( \beta \right)
\left[\begin{array}{c}
x(t) \\
y(t)
\end{array}\right] 
\label{Foucault_7} \, , \\
R_z\left( \beta \right) &=&
\left[\begin{array}{cc}
\cos\left( \beta\right) & \sin\left( \beta\right) \\
-\sin\left( \beta\right) & \cos\left( \beta\right) 
\end{array}\right]
\label{Foucault_8} \, .
\end{eqnarray}
We obtain in the rotated frame $(X,Y,z)$
\begin{eqnarray}
\ddot{X}(t) &=& - \omega^2_c X(t) + \Omega \dot{z}(t) 
\label{Foucault_9} \, , \\
\ddot{z}(t) &=& - \Delta^2 z(t) - \Omega \dot{X}(t) 
\label{Foucault_10} \, , \\
\ddot{Y}(t) &=& - \omega_c^2 Y(t)
\label{Foucault_11} \, . 
\end{eqnarray}
We see that the $Y(t)$-mode in Eq.~\ref{Foucault_11} decouples from the other $(X(t),z(t))$-modes.
The $Y(t)$-mode is thus a free running sinusoidal oscillation at frequency $\omega_c$.
In the following, we will focus on the case $\omega_c=\Delta$, which is the one studied in the paper, and for which simple analytical formulas can be derived.
We introduce the complex variable $\xi(t)=X(t)+iz(t)$, which after using Eqs.~\ref{Foucault_9},\ref{Foucault_10}, is shown to fulfill the simple differential equation
\begin{eqnarray}
\ddot{\xi}(t) &=& -i \Omega \dot{\xi}(t) - \omega^2_c \xi(t) 
\label{Foucault_12} \, . 
\end{eqnarray}
This equation can be solved exactly as
\begin{eqnarray}
\xi(t) &=& e^{-i \frac{\Omega}{2}t} \Big\lbrace 
\xi(0)\left\lbrack 
\cos\left( \tilde{\omega}_c t \right) 
+
i\frac{\Omega}{2\tilde{\omega}_c}\sin\left( \tilde{\omega}_c t \right) 
\right\rbrack 
\nonumber \\
&+& \frac{\dot{\xi}(0)}{\tilde{\omega}_c}\sin\left( \tilde{\omega}_c t \right) 
\Big\rbrace 
\label{Foucault_13} \, . 
\end{eqnarray}
with the renormalized cavity frequency $\tilde{\omega}_c$, provided by
\begin{eqnarray}
\tilde{\omega}_c &=& \sqrt{ \omega^2_c + \left( \frac{\Omega}{2} \right)^2 }
\label{Foucault_13_bis} \, . 
\end{eqnarray}
In the following, we use the initial conditions of the paper $x(0)=1$, $y(0)=z(0)=0$, $\dot{x}(0)=\dot{y}(0)=\dot{z}(0)$ and the choice $\alpha_u=0$ resulting in $\beta \approx 0$ for $\mathcal{R} \ll 1$.
From Eq.~\ref{Foucault_13}, we can find exact expressions for the rotated-variables $(X(t),z(t))$, and obtain finally an exact analytical solution for the trajectories in the non-rotated frame as
\begin{eqnarray}
x(t) &=& \cos\left( \frac{\Omega t}{2} \right) \cos\left( \tilde{\omega}_c t \right) 
+ \frac{\Omega}{2\tilde{\omega}_c} 
\sin\left( \frac{\Omega t}{2} \right) \sin\left( \tilde{\omega}_c t \right) 
\nonumber \, , \\
\label{Foucault_14} \\
y(t) &=& 0
\label{Foucault_15} \, , \\
z(t) &=& -\sin\left( \frac{\Omega t}{2} \right) \cos\left( \tilde{\omega}_c t \right) 
+ \frac{\Omega}{2\tilde{\omega}_c} 
\cos\left( \frac{\Omega t}{2} \right) \sin\left( \tilde{\omega}_c t \right) 
\nonumber \, . \\
\label{Foucault_16}  
\end{eqnarray}
We show in Fig.~\ref{fig:FigS7}, that this analytical expression generates trajectories (dashed-black curves) matching very well with the trajectories computed by integrating numerically Newton's equations (plain-red curves), and this, in the whole range of parameters investigated in the paper.
\subsubsection{Analogy with the Foucault pendulum ($\Omega \ll \omega_c$)}
\label{Weak_Coupling}
We then consider the regime of parameters for which $\Omega \ll \omega_c$, in which case $\tilde{\omega}_c \approx \omega_c$.
Neglecting the terms of order $\Omega/\omega_c$ in Eqs.~\ref{Foucault_14},\ref{Foucault_15}, the latter equations can by simplified to the approximate ones
\begin{eqnarray}
x(t) &\approx & \cos\left( \omega_c t \right) 
\cos\left( \frac{\Omega t}{2} \right)
\label{Foucault_17} \, , \\
y(t) &=& 0.0
\label{Foucault_18} \, , \\
z(t) &\approx & -\cos\left( \omega_c t \right) \sin\left( \frac{\Omega t}{2} \right)  
\label{Foucault_19} \, . 
\end{eqnarray}
Those equations describe beating oscillations in the $(x,z)$-plane at a fast frequency $\approx \omega_c$, together with a slow precession frequency $\Omega_P = \Omega/2$.
In this regime, both the equations of motion, and the obtained trajectories thus get equivalent to the ones describing the motion of a Foucault pendulum located at the North pole: in this analogy, the oscillation frequency of the pendulum is played by the cavity-frequency $\omega_c$, while its precession frequency due to the earth rotation becomes $\Omega_P$ in the effective model.
We get for the effective precession frequency the expression
\begin{eqnarray}
\Omega_P &=& \frac{\Omega_0}{2} \left( 1 + \Xi \right) 
\label{Foucault_20} \, , \\
\Omega_0 &=& \omega_P \left\lVert \vec{u}_\perp \right\rVert
\label{Foucault_20_bis} \, , \\
\Xi &=& \mathcal{R} \frac{\vec{e}_z\cdot\left( \vec{u} \times \vec{v} \right)}{\left\lVert \vec{u}_\perp\right\rVert^2} \label{Foucault_20_terce} \, . 
\end{eqnarray}
This reproduces well the oscillations of the trajectories (precession of the analogous Foucault pendulum) shown in Fig.~3-a) and Fig.~4 in the main paper.
After one precession, the anholonomy angle of the trajectory is perturbed by the gyrotropic coupling as $\Delta\Delta \theta \approx 2\pi\Xi$, recovering the expression of the paper. 
However, we note that the approximate Eqs.~\ref{Foucault_17},\ref{Foucault_18},\ref{Foucault_19}, do not capture well the presence of cusp singular points in the trajectories (see Fig.~\ref{fig:FigS7}).
To obtain them, one needs to include in the calculation the next non-vanishing terms of order $\Omega/2\omega_c$ in Eqs.~\ref{Foucault_14},\ref{Foucault_15},\ref{Foucault_16} (see next Sec.~\ref{Strong_Coupling}).
\subsubsection{General case of USC ($\Omega \geq \omega_c$)}
\label{Strong_Coupling}
%
%
\begin{figure}[tbh]
\includegraphics[width=1.0\linewidth, angle=-0]{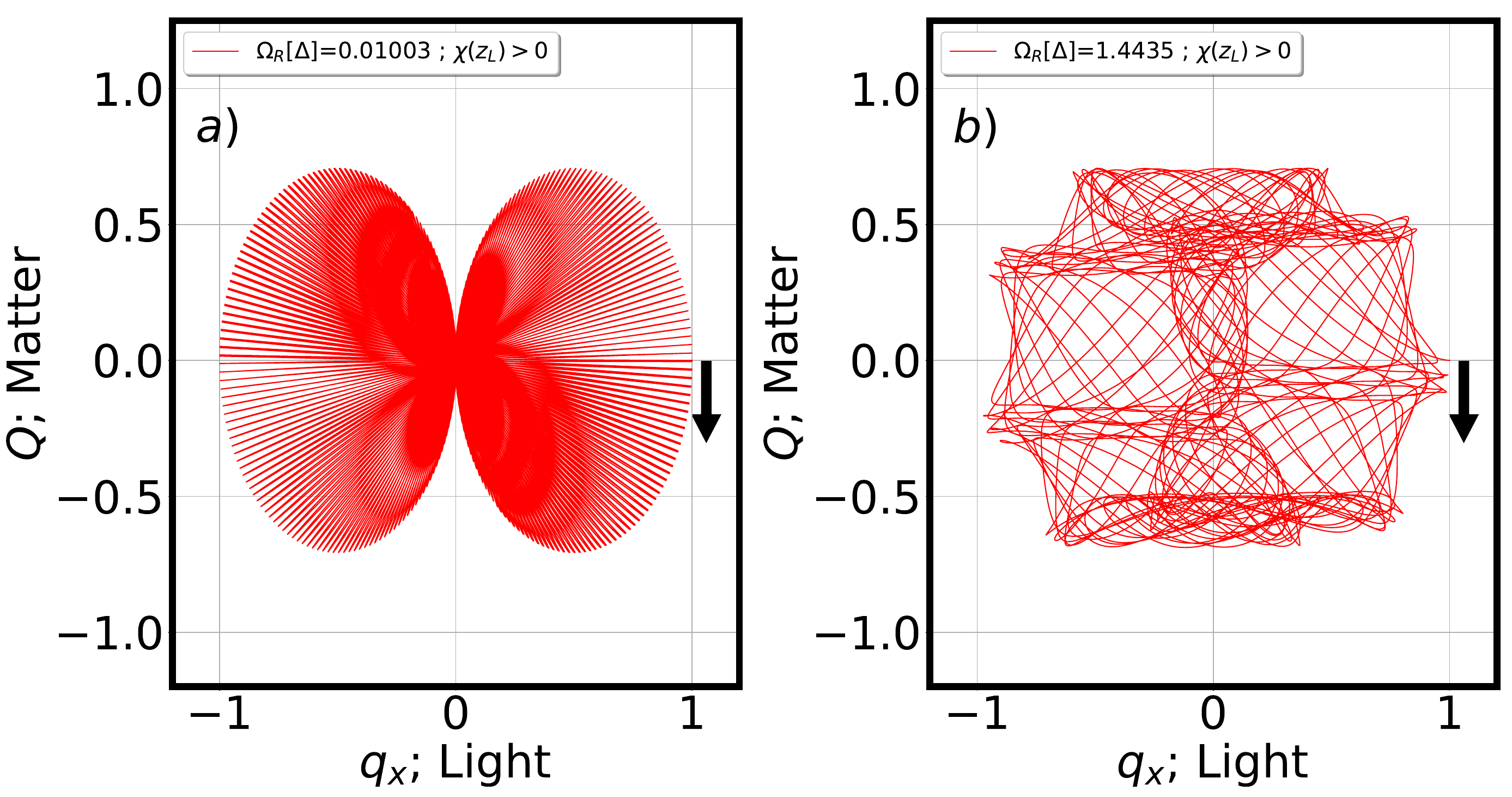}
\caption{
a) Plot of the classical trajectory $\left(q_x(t),q_y(t),Q(t)\right)$ in the plane $\left(q_x,Q\right)$ with $\Omega_R\approx 0.01$, during the time-interval $t \in \left\lbrack 0, 1200/\Delta \right\rbrack$.
b) Same plot, but with $\Omega_R\approx 1.44$, during the time-interval $t \in \left\lbrack 0, 400/\Delta \right\rbrack$.
Parameters are: $\Delta=1.0$, $\omega_c=1.0$, $m=1$, $\alpha_u=\frac{\pi}{4}$, $\alpha=\frac{\pi}{2}$, $z_L=\frac{L}{4}$, and $\chi(z_L)=\mathcal{R}=10^{-3}>0$.
Initial conditions: $q_x(0)=1.0$, $q_y(0)=Q(0)=0.0$, $\dot{q}_x(0)=\dot{q}_y(0)=\dot{Q}(0)=0.0$.
\label{fig:FigS4}}
\end{figure}
At higher cVRS ($\Omega_R \geq \omega_c$), the analogy with a Foucault pendulum breaks down, and the classical dynamics of the fictive particle exhibits cyclotron gyrations around the direction of the effective $\vec{B}$-field written in Eq.~\ref{Classical_Model_dim_3_Phy_Int2}.
We thus come back to the analysis of exact Eqs.~\ref{Foucault_14},\ref{Foucault_15},\ref{Foucault_16}, which can be rewritten in the following canonical form
\begin{eqnarray}
x(\theta) &=& r \left\lbrace
\left( q-1 \right) \cos\left(\theta\right)
+
\cos\left\lbrack\left( q-1 \right)\theta\right\rbrack
\right\rbrace 
\label{Strong_Coupling_1} \, , \\
y(\theta) &=& 0.0
\label{Strong_Coupling_2} \, , \\
z(\theta) &=& r \left\lbrace
\left( q-1 \right) \sin\left(\theta\right)
-
\sin\left\lbrack\left( q-1 \right)\theta\right\rbrack
\right\rbrace 
\label{Strong_Coupling_3} \, , 
\end{eqnarray}
where we introduced the $(\theta,r,q)$ parameters
\begin{eqnarray}
\theta &=& \left( \tilde{\omega}_c - \frac{\Omega}{2} \right) t 
\label{Strong_Coupling_4} \, , \\
r &=& \frac{1}{2}\left\lbrace 1 - \frac{\Omega}{2\tilde{\omega}_c} \right\rbrace
\label{Strong_Coupling_5} \, , \\
q &=& \frac{2}{1 - \frac{\Omega}{2\tilde{\omega}_c}}
\label{Strong_Coupling_6} \, .
\end{eqnarray}
%
%
%
\begin{figure}[tb]
\includegraphics[width=1.0\linewidth, angle=-0]{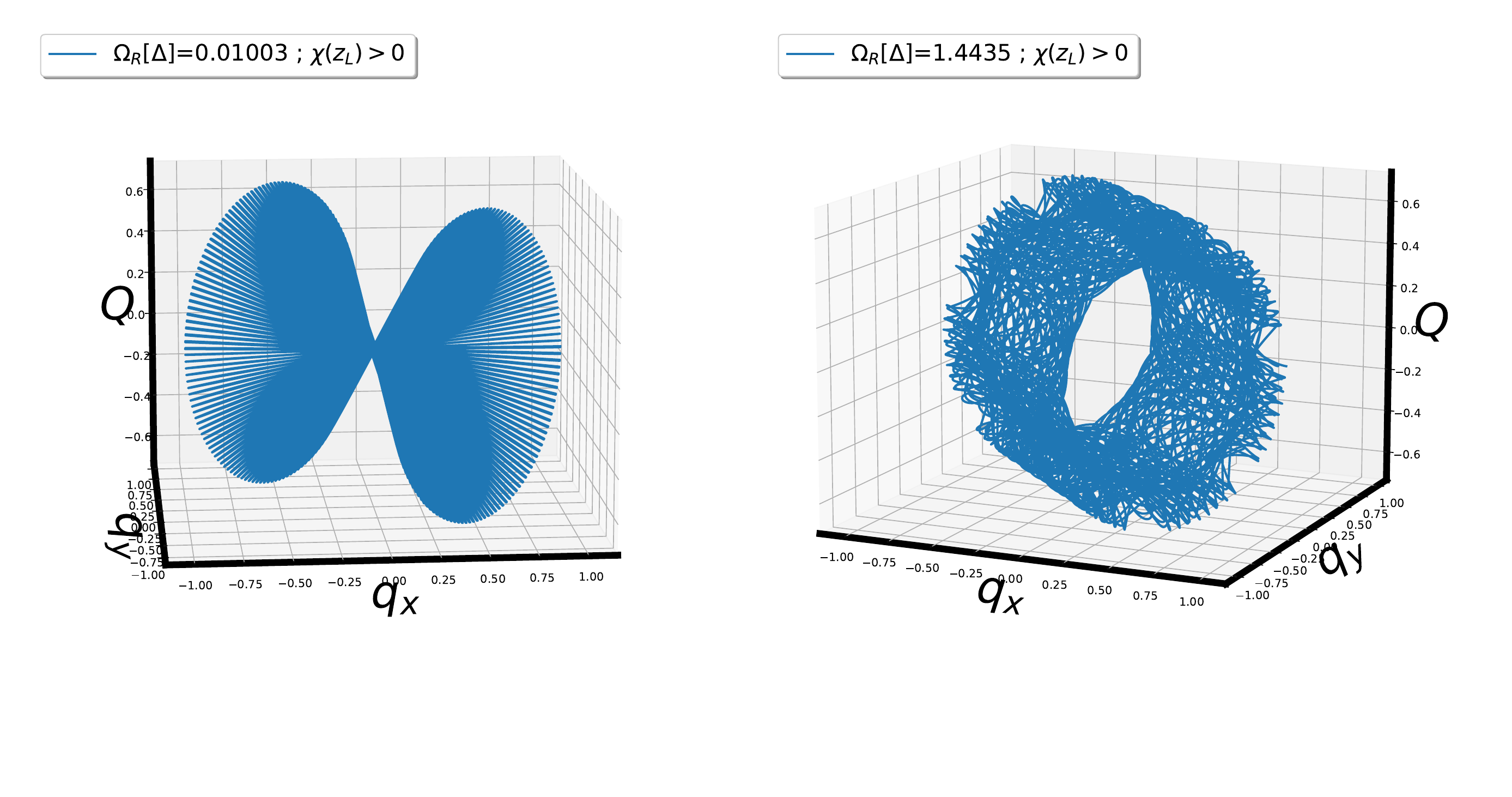}
\caption{
3D-plot of the classical trajectory $\left(q_x(t),q_y(t),Q(t)\right)$ with the same parameters as in Fig.~\ref{fig:FigS4}.
\label{fig:FigS5}}
\end{figure}
\textit{In this form, Eqs.\ref{Strong_Coupling_1},\ref{Strong_Coupling_3} are particularly transparent 
and provide an accurate description of the classical trajectories in the full range of parameters investigated in the paper. 
They describe an hypocycloid trajectory \footnote{See for instance \url{https://mathcurve.com/courbes2d/hypocycloid/hypocycloid.shtml} for graphical illustrations and selected animations of hypocycloid curves, with some of their mathematical properties} in the $(x,z)$-plane}, with a radius $R=1$ of the external fixed-circle provided by the choice of initial conditions and conservation of mechanical energy (the $\vec{B}$-field does not work), and a radius $r=R/q$ of the internal moving-circle. 
The angle $\theta$ describes an oscillation at angular frequency $\tilde{\omega}_c - \frac{\Omega}{2}$, while the parameter $q=R/r$ is the ratio between the radius of the fixed-circle and the moving-circle. 
\textit{The hypocycloid trajectory of the fictive mobile, thus takes place inside a ring-manifold} with big-circle of radius $R$ and low-circle of radius $R_{\rm{min}}=R-2r$ given by
\begin{eqnarray}
R_{\rm{min}} &=& \frac{\Omega}{2\tilde{\omega}_c}
\label{Strong_Coupling_7} \, .
\end{eqnarray}
This dimensionless parameter contains both effects of the light-matter coupling-strength and the chiral-coupling encoded in Eq.~\ref{Foucault_20}. 
In the weak-coupling regime ($\Omega  \ll \omega_c$), the lower-circle of the ring has a vanishing radius $R_{\rm{min}} \approx 0$ and thus converges to a single point at the center of the fixed-circle, recovering the trajectory of a precessing Foucault pendulum as described in Sec.\ref{Weak_Coupling} and shown in Fig.~\ref{fig:FigS7}-a).
In the deep-USC regime ($\Omega  \gg \omega_c$ and $\tilde{\omega}_c\approx \Omega/2$), the lower-circle of the ring gets close to the fixed-circle with a radius $R_{\rm{min}} \approx 1$: the mobile thereby undergoes cyclotron gyrations around the direction of the effective $\vec{B}$-field, with a radial velocity $\approx \Omega/2$.
In the USC regime ($\Omega  \geq \omega_c$), the trajectories takes the form of developed hypocycloids as shown in Fig.~\ref{fig:FigS7}-b).
It is interesting to notice that the cusp singularities of the hypocycloid curves are achieved at times $t_k \geq 0$ given by the expression
\begin{eqnarray}
t_k &=& \frac{k}{q}\frac{2\pi}{\tilde{\omega}_c - \frac{\Omega}{2}}; \mbox{ } k \in \mathbb{N}  
\label{Strong_Coupling_7} \, .
\end{eqnarray}
\textit{The trajectory is thus closed, namely it comes back to its initial point, if and only if 
$q$ is a rational number} ($q \in \mathbb{Q}$ with $q \ge 2$).
The latter condition imposes the following constraint on $\Omega/\omega_c$
\begin{eqnarray}
\frac{\Omega}{2\omega_c} &=& \frac{q-2}{2\sqrt{q-1}}; \mbox{ } q \in \mathbb{Q} \cap \left\lbrack 2, +\infty \right\lbrack  
\label{Strong_Coupling_7_bis} \, . 
\end{eqnarray}
\textit{If $q$ is an irrational number, the trajectories are not closed and fill the ring manifold in a dense-way (quasi-periodic nature of the trajectories \cite{arnol2013mathematical})}.
\subsubsection{Changing the $\alpha_u$-parameter}
\label{Initial_Param}
To illustrate the complexity hidden in the classical dynamics of the analogous model developed in Sec.~\ref{Classical_Model_dim_3_Phy_Int}, we show in Fig.~\ref{fig:FigS4} the classical trajectories obtained by numerical integration of Newton Eq.~\ref{Classical_Model_dim_3_Phy_Int2}, but with a choice of parameter $\alpha_u=\pi/4$ differing from the one chosen in Fig.~3 of the paper (for which $\alpha_u=0$).
Interestingly, the polaritonic spectra (as shown in Fig.~2 of the paper) are the same for either choice of $\alpha_u=0$ or $\alpha_u=\pi/4$, since they depend only on the gyrotropic coupling parameter $\chi(z_L)\propto \mathcal{R}\vec{e}_z \cdot\left( \vec{u} \times \vec{v} \right)$ (see Eq.~\ref{ChiralClassicalHbos2Modes20}), which is the same in both cases.
However, as shown in Fig.~\ref{fig:FigS4}, the obtained classical trajectories are completely different with a different choice of $\alpha_u$-parameter (and thus a different value of $\beta$), despite the same initial value of the coordinates and velocities.
The trajectories are now developing in 3D, where they span some 2D-manifolds shown in Fig.~\ref{fig:FigS5}.
Projecting the coordinates on a plane orthogonal to the new direction of the effective $\alpha_u$-dependent $\vec{B}$-field, we recover 2D-trajectories that are similar to the ones shown in the paper (for $\alpha_u=0$). 
\textit{This means that the notion of change in the anholonomy angle due to the gyrotropic coupling is a robust and central feature characterizing the chiral light-matter interaction mechanism.}
It would also be of interest to probe the connection between this analysis of the topology of the trajectories in the analogous classical model, with a deeper analysis of the information encoded into the phase of the wave-function for the quantum model \cite{hannay_angle_1985,berry_geometric_1988}. 
%

\section{3D-bulk configuration of molecules in the optical cavity}
\label{3DChiralityorderedHopfield}
\subsection{Hopfield polaritons}
\label{HopfieldPolChiral3D}
In this section, we investigate briefly the problem of a 3D-bulk configuration of coupled molecules inside the optical cavity.
In this case, the coupled molecular aggregate fills homogeneously the whole cavity volume.
We suppose that the molecules are identical, with the same orientation of their transition-dipoles, and that each molecule in the Fabry-Pérot sample is located at the vertices of a cubic lattice of lattice parameter $a$, at a position labelled by the list of integers $j=(n_x,n_y,p)$.
The total number of molecules is $N=N_xN_yN_z$, with $N_{\alpha}$ the number of molecules in  each $\alpha=x,y,z$-direction.
We follow tightly the approach developed in Sec.\ref{2DChiralityorderedHopfield}, and generalize the expression of the bosonized Hamiltonian written in Eq.~\ref{ChiralHbos2Modes1} to this 3D-bulk configuration
\begin{eqnarray}
H^{(\rm{3D})}_{\rm{bos}} &=& H_c + H_e + H_{D} + H_{M} + H_{SD}
\label{ChiralHbos2Modes3D1} \, , \\
H_c &=& \hbar\omega_c \left\lbrace 
a_{x}^\dagger a_{x} + a_{y}^\dagger a_{y} + 1 
\right\rbrace
\label{ChiralHbos2Modes3D2} \, , \\
H_e &=& \hbar \Delta \sum_{j=1}^N \left\lbrace b_{j}^\dagger b_{j} + \frac{1}{2} \right\rbrace
\label{ChiralHbos2Modes3D3} \, , \\
H_D &=& -i\hbar \Omega_R \left( B_u + B_u^\dagger \right)
\Big\lbrace
u_{x} \left( a_{x} - a_{x}^\dagger \right)
\nonumber \\
&+& u_{y}
\left( a_{y} - a_{y}^\dagger \right)
\Big\rbrace 
\label{ChiralHbos2Modes3D4} \, , \\
H_M &=& -i\hbar \mathcal{R} \Omega_R \left( B_v - B_v^\dagger \right) \Big\lbrace
v_{x} 
\left( a_{y} + a_{y}^\dagger \right)
\nonumber \\
&-& v_{y} 
\left( a_{x} + a_{x}^\dagger \right)
\Big\rbrace 
\label{ChiralHbos2Modes3D5} \, ,  \\
H_{SD} &=& \hbar D
\left( B_u + B_u^\dagger \right)^2
\label{ChiralHbos2Modes3D6} \, ,
\end{eqnarray}
where we introduced 
\begin{eqnarray}
\Omega_R &=& g \sqrt{\frac{N}{2}}
\label{3DHop1} \, , \\
D &=& \frac{\Omega_R^2}{\omega_c} \left\lVert \vec{u}_\perp \right\rVert^2 
\label{3DHop2} \, ,
\end{eqnarray}
Note that in comparison to Eq.~\ref{2DHop2}, \textit{the collective Rabi-splitting $\Omega_R$ is divided by a factor $\sqrt{2}$, coming from the quadratic integration of the cVRS along the $z$ optical-axis of the cavity.}
We also introduced two (instead of one) independent collective excitonic variables
\begin{eqnarray}
B_u &=& \sqrt{\frac{2}{N_z}} \sum_{p=0}^{N_z-1} \sin\left(\theta_{z_p}\right) B_p
\label{3DHop3} \, , \\
B_v &=& \sqrt{\frac{2}{N_z}} \sum_{p=0}^{N_z-1} \cos\left(\theta_{z_p}\right) B_p
\label{3DHop4} \, , 
\end{eqnarray}
with 
\begin{eqnarray}
B_p &=& \frac{1}{\sqrt{N_xN_y}}\sum_{n_x=0}^{N_x-1}\sum_{n_y=0}^{N_y-1}
b_{(n_x,n_y,p)}
\label{3DHop5} \, .
\end{eqnarray}
In Eq.~\ref{3DHop5}, $B_p$ is the Fourier-transform of the microscopic exciton destruction operators $b_j \equiv b_{(n_x,n_y,p)}$, in the transverse (Oxy) plane (located at $z=z_p$), and at wave-vector $\vec{k}_\perp=\vec{0}$.
The $(B_u,B_v)$ operators are quasi-bosonic destruction operators obtained after spatial integration along the $z$ optical-axis.
They fulfil the commutation relations, in the harmonic approximation of the Hopfield model \cite{hopfield_theory_1958}
\begin{eqnarray}
\left\lbrack B_u, B_u^\dagger \right\rbrack &=& \left\lbrack B_v, B_v^\dagger \right\rbrack = 1
\label{3DHop6} \, , \\
\left\lbrack B_u, B_v^\dagger \right\rbrack &=& \left\lbrack B_u, B_v \right\rbrack = 0
\label{3DHop7} \, .
\end{eqnarray}
As in the previous Sec.~\ref{HopfieldPolChiral}, we perform a Bogoliubov transformation of the quadratic Hamiltonian $H^{(\rm{3D})}_{\rm{bos}}$, obtaining $\chi$-operators that follow the same linear eigenvalue problem as in Eqs.\ref{Bog1},\ref{Bog2}, but with a different $\underline{\underline{g}}$-metric 
\begin{eqnarray}
\underline{\underline{g}} &=& \mbox{diag} \left( 1,1,1,1,-1,-1,-1,-1 \right)
\label{3DHop7_bis}  \, ,
\end{eqnarray}
and a larger number of components taken into account in the linear combination of operators
\begin{eqnarray}
\chi &=& \underline{\hat{A}}^t \cdot \underline{C}
\label{3DHop8} \, , \\
\underline{\hat{A}} &=& \left\lbrack
a_x, a_y, B_u, B_v, a^\dagger_x, a^\dagger_y, B_u^\dagger, B_v^\dagger
\right\rbrack
\label{3DHop9} \, , \\
\underline{C} &=& \left\lbrack
\alpha_x, \alpha_y, \gamma_u, \gamma_v, \beta_x, \beta_y, \delta_u, \delta_v
\right\rbrack
\label{3DHop10} \, . 
\end{eqnarray}
We obtain a Hopfield-matrix for the 3D-bulk problem of dimension $8\times 8$  (similar to, but larger than Eq.~\ref{2DHop8})
%
\begin{widetext}
\begin{eqnarray}
\underline{\underline{\mathcal{H}}}^{(\rm{3D})} &=&
\left[\begin{array}{cccccccc}
\omega_c & 0 & -i\Omega_R u_x & -i\Omega_R \mathcal{R}v_y & 0 & 0 & i\Omega_R u_x & -i\Omega_R \mathcal{R}v_y \\
0 & \omega_c & -i\Omega_R u_y & i\Omega_R \mathcal{R}v_x & 0 & 0 & i\Omega_R u_y & i\Omega_R \mathcal{R}v_x \\
i\Omega_R u_x & i\Omega_R u_y  & \Delta + 2D & 0 & i\Omega_R u_x & i\Omega_R u_y  & -2D & 0 \\
i\Omega_R \mathcal{R} v_y & -i\Omega_R \mathcal{R} v_x  & 0 & \Delta & -i\Omega_R \mathcal{R} v_y  & i\Omega_R \mathcal{R} v_x & 0 & 0 \\
0 & 0 & i\Omega_R u_x & -i\Omega_R \mathcal{R}v_y & -\omega_c & 0 & -i\Omega_R u_x  & -i\Omega_R \mathcal{R}v_y  \\
0 & 0 & i\Omega_R u_y & i\Omega_R \mathcal{R}v_x & 0 & -\omega_c & -i\Omega_R u_y & i\Omega_R \mathcal{R}v_x  \\
i\Omega_R u_x & i\Omega_R u_y & 2D & 0 & i\Omega_R u_x & i\Omega_R u_y & -\left( \Delta + 2D \right) & 0 \\
-i\Omega_R \mathcal{R} v_y & i\Omega_R \mathcal{R} v_x & 0 & 0 & i\Omega_R \mathcal{R} v_y & -i\Omega_R \mathcal{R} v_x & 0 &  -\Delta  
\end{array}\right]
\nonumber \, . \\
\label{3DHop11}
\end{eqnarray}
\end{widetext}
%
The characteristic polynomial $P^{(\rm{3D})}(\lambda)$ of this Hopfield-matrix, can still be obtained analytically as
\begin{eqnarray}
&&P^{(\rm{3D})}(\lambda) = \left( \lambda^2-\omega_c^2 \right)^2\left( \lambda^2-\Delta^2 \right)^2
-
4  \frac{\Omega_R^2}{\omega_c} \Delta \Big\lbrace
\nonumber \\ 
&&\lambda^2 \left\lVert \vec{u}_\perp \right\rVert^2 
+ \mathcal{R}^2 \omega_c^2 \left\lVert \vec{v}_\perp \right\rVert^2
\Big\rbrace
\left( \lambda^2-\omega_c^2 \right) \left( \lambda^2-\Delta^2 \right)
\nonumber \\
&+& 16 \mathcal{R}^2 \lambda^2 \Omega_R^4 \Delta^2
\left( \vec{u}_\perp \cdot \vec{v}_\perp \right)^2
\label{3DHop12}  \, .
\end{eqnarray}
This polynomial has 4 positive roots, corresponding to the collective polaritonic excitations in the 3D-bulk configuration: one lower-polariton branch (labelled $\omega^{\rm{3D}}_-$), one upper-polariton (labelled $\omega^{\rm{3D}}_+$), and two middle-polaritons (labelled with the same $\omega^{\rm{3D}}_0$).
\subsection{Analogous classical model in dimension $3+1$}
\label{Classical_Model_dim_4}
%
%
\begin{figure}[!tbh]
\centering
\includegraphics[width=1.0\linewidth]{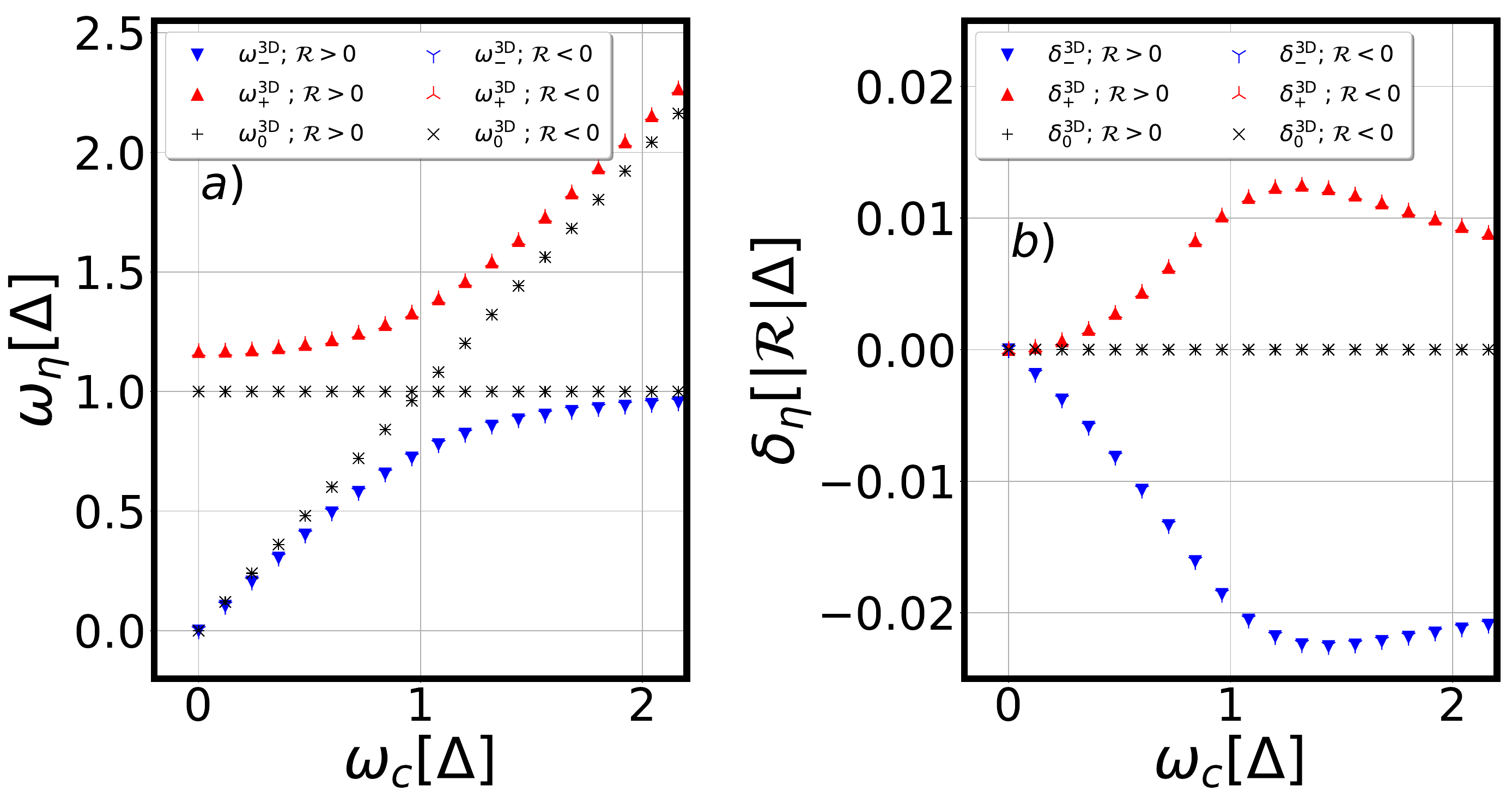}
\caption{
a) Numerical spectra of the Hopfield polaritons obtained from the 3D-bulk Hopfield-matrix in Eq.~\ref{3DHop11}, as a function of the cavity frequency $\omega_c$.
b) Same curves but for the relative shift of the polaritonic spectra $\delta^{\rm{3D}}_\eta = \omega^{\rm{3D}}_{\eta;\mathcal{R}} - \omega^{\rm{3D}}_{\eta;\mathcal{R}=0}$ (in units of $\left\lvert \mathcal{R} \right\rvert\Delta$).
Parameters are: $\Delta=1.0$, $m=1$, $z_L=\frac{L}{4}$, $\Omega^{\rm{res}}_R=0.3$, $\alpha_u=\frac{\pi}{4}$, $\alpha=\frac{\pi}{2}$, and $\chi({z_L})=\mathcal{R}=0.1$.
}
\label{fig:FigS6}
\end{figure}
In close analogy to Sec.\ref{Classical_Model_dim_3}, we obtain the classical limit of the bosonized Hamiltonian $H ^{(\rm{3D})}_{\rm{bos}}$ written in Eq.~\ref{ChiralHbos2Modes3D1}, as 
\begin{eqnarray}
H^{(\rm{3D,cl})}_{\rm{bos}} &=& \frac{\left\lVert \vec{p}_\perp \right\rVert^2}{2}
+ \frac{\omega^2_c}{2}\left\lVert \vec{q}_\perp \right\rVert^2
+ \frac{P_u^2}{2}
+ \frac{\Delta^2}{2} Q_u^2
\nonumber \\
&+& \frac{P_v^2}{2}
+ \frac{\Delta^2}{2} Q_v^2
+
\omega_P \vec{u}_\perp \cdot \vec{p}_\perp Q_u 
\nonumber \\
&+&\left( \vec{\Omega}_g \times \vec{v}_\perp \right) \cdot \vec{q}_\perp P_v
+
\frac{\omega^2_P}{2} \left\lVert \vec{u}_\perp \right\rVert^2 Q_u^2
\nonumber \, , \\
\label{ChiralClassicalHbos2Modes3D1}  
\end{eqnarray}
with
\begin{eqnarray}
\omega_P &=& \sqrt{\frac{2N\Delta}{\omega_c}} g
\label{ChiralClassicalHbos2Modes3D2} \, , \\
\vec{\Omega}_g &=& \Omega_g \vec{e}_z = \mathcal{R}\frac{\omega_c}{\Delta} \omega_P \vec{e}_z 
\label{ChiralClassicalHbos2Modes3D3} \, . 
\end{eqnarray}
When the coordinates $(q_x,p_x, q_y, p_y, Q_u, P_u, Q_v, P_v)$ are considered as classically conjugated variables, we can derive classical Hamilton equations from Eq.~\ref{ChiralClassicalHbos2Modes3D1}, from which classical Newton equations for the 4 coordinates $(q_x,q_y, Q_u, Q_v)$ alone are obtained
\begin{eqnarray}
\ddot{Q}_u &=& - \Delta^2 Q_u  + \left(\vec{\Omega}_g \times \vec{v}_\perp \right)\cdot \dot{\vec{q}}_\perp
\label{ChiralClassicalHbos2Modes3D4} \, , \\
\ddot{Q}_v &=& - \Delta^2 Q_v - \omega_P \vec{u}_\perp \cdot \dot{\vec{q}}_\perp
\label{ChiralClassicalHbos2Modes3D4} \, , \\
\ddot{\vec{q}}_\perp &=& -\omega^2_c \vec{q}_\perp
+
\left(\vec{\Omega}_g \times \vec{v}_\perp \right)
\left\lbrack
\left(\vec{\Omega}_g \times \vec{v}_\perp \right)\cdot\vec{q}_\perp
\right\rbrack
\nonumber \\
&+&
\omega_P \vec{u}_\perp \dot{Q}_v -\left(\vec{\Omega}_g \times \vec{v}_\perp \right)\dot{Q}_u
\label{ChiralClassicalHbos2Modes3D5} \, . 
\end{eqnarray}
\textit{Those equations describe the classical motion of a fictive mobile of unit mass in dimension 3+1.}
Non-trivial eigensolutions of this system of linear equations in Fourier, are found as zeros of the secular determinant $\Delta^{(\rm{3D,cl})}(\omega)$
\begin{eqnarray}
&&\Delta^{(\rm{3D,cl})}(\omega) = \left( \omega^2-\omega_c^2 \right)^2\left( \omega^2-\Delta^2 \right)^2
\nonumber \\
&-&
\left( \omega^2-\omega_c^2 \right)\left( \omega^2-\Delta^2 \right)
\Big\lbrace
\Delta^2 \Omega_g^2\left\lVert \vec{v}_\perp \right\rVert^2
+
\omega^2 \omega_P^2\left\lVert \vec{u}_\perp \right\rVert^2
\Big\rbrace
\nonumber \\
&+&
\omega^2 \Delta^2 \Omega_g^2 \omega_P^2
\left( \vec{u}_\perp \cdot \vec{v}_\perp \right)^2 
\label{ChiralClassicalHbos2Modes3D6}  \, . 
\end{eqnarray}
Using Eqs.~\ref{ChiralClassicalHbos2Modes3D2},\ref{ChiralClassicalHbos2Modes3D3}, it is straightforward to show that the determinant $\Delta^{(\rm{3D,cl})}(\omega)$ in Eq.~\ref{ChiralClassicalHbos2Modes3D6} is the same as the characteristic polynomial $P^{(\rm{3D})}(\omega)$ written in Eq.~\ref{3DHop12} for the Hopfield problem.
Thus both problems have the same eigensolutions and the same collective polariton excitations. 
Moreover, we see that both polynomials do not contain any term linear in $\mathcal{R}$: the only terms containing the gyrotropic coupling are those of order $\mathcal{R}^2$. 
\textit{The corresponding lower and upper polariton spectra in the 3D-bulk configuration are thus functions of $\mathcal{R}^2$ and thus, do not depend anymore on the enantiomeric class of the active molecules, nor on the sign of the scalar quantity $\mathcal{R}\vec{e}_z\cdot\left( \vec{u} \times \vec{v} \right)$.}
This is in contrast to the 2D-layer case (see Eqs.~\ref{2DHop9},\ref{ChiralClassicalHbos2Modes19}):
in the case of the ordered 3D-bulk configuration, \textit{the effect of the gyrotropic coupling on the differential polariton spectra thus does not survive to the $z$-integration along the optical-axis performed in Eqs.~\ref{3DHop3},\ref{3DHop4}.}
For completeness, we show in Fig.~\ref{fig:FigS6}, the polaritonic spectra $\omega^{\rm{3D}}_\eta$ and differential $\delta^{\rm{3D}}_\eta$-signals, computed after numerical diagonalization of the $\underline{\underline{\mathcal{H}}}^{(\rm{3D})}$ Hopfield-matrix written in Eq.~\ref{3DHop11}.
The lower (upper) polariton $\omega^{\rm{3D}}_\eta$ is labelled with the index $\eta = -(+)$ and appears with blue(red) points in the figure, while the remaining two middle polaritons $\omega^{\rm{3D}}_0$ are both labelled with $\eta=0$ and both appear as black points for simplicity. 
The parameters are chosen to correspond to those of Fig.~\ref{fig:FigS1}, but with the choice of a large parameter $\mathcal{R}=0.1$.
\textit{In contrast to the 2D-layer case, we confirm in Fig.~\ref{fig:FigS6}-b), the absence of differential-signal due to the gyrotropic coupling, namely that
$\delta^{\rm{3D}}_\eta(\mathcal{R})=\delta^{\rm{3D}}_\eta(-\mathcal{R})$ (the $\delta^{\rm{3D}}_\eta$-signals do not reverse sign upon reversal of the sign of $\mathcal{R}$).}

\bibliography{biblio_SupMat}